\documentclass{LMCS}

\def\doi{8(3:16)2012}
\lmcsheading%
{\doi}
{1--29}
{}
{}
{Oct.~15, 2011}
{Sep.~14, 2012}
{}

\usepackage{mathpartir}
\usepackage{alltt}
\usepackage{amssymb}
\usepackage{amsmath}
\usepackage{rotating}
\usepackage{url}
\usepackage{mathtools}
\usepackage{hyperref}
\usepackage{enumerate}

\theoremstyle{plain}\newtheorem{axiom}[thm]{Axiom}

\newcommand{\CA}{\mathcal{A}}
\newcommand{\CC}{\mathcal{C}}
\newcommand{\CE}{\mathcal{E}}
\newcommand{\CP}{\mathcal{P}}
\newcommand{\R}{\ensuremath{\mathcal{R}}}
\newcommand{\CS}{\mathcal{S}}
\newcommand{\CT}{\mathcal{T}}
\newcommand{\CX}{\mathcal{X}}

\newcommand{\pNat}{\ensuremath{\mathbb{N}}}

\newcommand{\X}{\ensuremath{\mathsf{X}}}

\newcommand{\LA}{\ensuremath{\mathsf{LA}}}
\newcommand{\AC}{\ensuremath{\mathsf{AC}}}

\newcommand{\ACX}{\textsf{AC(\X)}}
\newcommand{\acEmpty}{\textsf{AC($\emptyset$)}}
\newcommand{\acLAMacro}{\textsf{AC(\LA)}}

\newcommand{\TF}{{\ensuremath{{\mathcal T}_{\Sigma}}}}
\newcommand{\TFX}{{\ensuremath{{\mathcal T}_{\Sigma}(\mathcal X)}}}

\newcommand{\nf}[2]{\ensuremath{#2\!\downarrow_{#1}}}
\newcommand{\make}[1]{\ensuremath{[\![ #1 ]\!]}}
\newcommand{\can}{{\tt can}}
\newcommand{\canX}{{\tt can}_{\X}}
\newcommand{\canAC}{{\tt can}_{AC}}

\newcommand{\multiset}[1]{\ensuremath{\{\!\!\{#1\}\!\!\}}}

\newcommand{\ergo}  {\textsc{Alt-Ergo}}
\newcommand{\trivial}{$\mathbf{Tri}$vial}
\newcommand{\orient}{$\mathbf{Ori}$ent}
\newcommand{\simplify}{$\mathbf{Sim}$plify}
\newcommand{\compose}{$\mathbf{Com}$pose}
\newcommand{\collapse}{$\mathbf{Col}$lapse}
\newcommand{\deduce}{$\mathbf{Ded}$uce}
\newcommand{\xsolve}{$\mathbf{Ori}$ent}
\newcommand{\bottom}{$\mathbf{Bot}$tom}

\newcommand{\orientbis}{$\mathbf{Ori}$}
\newcommand{\simplifybis}{$\mathbf{Sim}$}
\newcommand{\composebis}{$\mathbf{Com}$}
\newcommand{\collapsebis}{$\mathbf{Col}$}
\newcommand{\deducebis}{$\mathbf{Ded}$}
\newcommand{\xsolvebis}{$\mathbf{Ori}$}

\newcommand{\fsolveX}{\fsolve_{\,\X}}
\newcommand{\fheadcp}{\ensuremath{\mathtt{headCP}}}
\newcommand{\fsolve}{\ensuremath{\mathtt{solve}}}

\newcommand{\OM}{\textsc{om}}
\newcommand{\TO}{\textsc{to}}

\newcommand{\rsa}{\rightsquigarrow}

\newcommand{\downsquigarrow}
{\ensuremath{\raisebox{2.5mm}{\begin{rotate}{270}$\rsa$\end{rotate}}~~}}
\newcommand{\red}[1]{\ensuremath{Red(#1)}}

\newcommand{\leftsquigarrow}
{\ensuremath{\mbox{~~~~\hspace{2ex}}\raisebox{2mm}{\begin{rotate}{180}$\rsa$\end{rotate}}}}

\newcommand{\compl}[2]{
  \ensuremath{
     \langle
      ~~~ #1 ~~~ | 
      ~~~ #2 ~~~
    \rangle  }} 

\newcommand{\complbis}[2]{
  \ensuremath{
     \langle
      ~ #1 ~ | 
      ~ #2 ~ 
    \rangle  }} 

\begin{document}

\title[Canonized Rewriting and Ground AC Completion Modulo Shostak
  Theories]{Canonized Rewriting and Ground AC Completion Modulo Shostak
  Theories : Design and Implementation}

\author[S.~Conchon]{Sylvain Conchon}
\address{LRI, Univ Paris-Sud, CNRS, Orsay F-91405\\
INRIA Saclay -- Ile-de-France, ProVal, Orsay, F-91893}
\email{\{Sylvain.Conchon, Evelyne.Contejean, Mohamed.Iguernelala\}@lri.fr}
\thanks{Work partially supported by the French ANR project ANR-08-005 Decert.}

\author[\'E.~Contejean]{\'Evelyne Contejean}

\author[M.~Iguernelala]{Mohamed Iguernelala}


\begin{abstract}
  AC-completion efficiently handles equality modulo associative and
  commutative function symbols. When the input is ground, the
  procedure terminates and provides a decision algorithm for the word
  problem. In this paper, we present a modular extension of ground
  AC-completion for deciding formulas in the combination of the theory
  of equality with user-defined AC symbols, uninterpreted symbols and
  an arbitrary signature disjoint Shostak theory $\mathsf{X}$.  Our algorithm,
  called \textsf{AC($\mathsf{X}$)}, is obtained by augmenting in a modular way
  ground AC-completion with the canonizer and solver present for the
  theory  $\mathsf{X}$. This integration rests on canonized rewriting, a new
  relation reminiscent to normalized rewriting, which integrates
  canonizers in rewriting steps. \textsf{AC($\mathsf{X}$)} is proved sound, complete
  and terminating, and is implemented to extend the core of the
  \textsc{Alt-Ergo} theorem prover.
\end{abstract}

\maketitle

\section{Introduction}

The mechanization of mathematical proofs is a research domain that
receives an increasing interest among mathematicians and computer
scientists. In particular, automated theorem provers (ATP) are now
used in several contexts (\emph{e.g.} proof of programs, interactive
provers) to prove ``simple'' but overwhelming intermediate results.
While more and more efficient, ATP have difficulties to handle some
mathematical operators, such as union and intersection of sets, which
satisfy the following associativity and commutativity (AC) axioms
\[
\begin{array}{rclr}
\forall x. \forall y. \forall z.~ u(x,u(y,z))& \,=\, &u(u(x,y),z)
& \textmd{(A)}\\ 
\forall x. \forall y.~ u(x,y)& \,=\, &u(y,x) & \qquad\textmd{(C)}
\end{array}
\]
%
Indeed, the mere addition of AC axioms to a prover will usually glut
it with plenty of useless equalities which will strongly impact its
performances\footnote{Given a term $t$ of the form
  $u(c_1,u(c_2,\ldots,u(c_n,c_{n+1})\ldots)$, the axiomatic approach
  may have to explicitly handle the $(2n)!/n!$ terms equivalent to
  $t$.}. In order to avoid this drawback, built-in procedures have
been designed to efficiently handle AC symbols.  For instance,
SMT-solvers incorporate dedicated decision procedures for some
\emph{specific} AC symbols such as arithmetic or boolean operators. On
the contrary, algorithms found in resolution-based provers such as
AC-completion allow a powerful \emph{generic} treatment of
user-defined AC symbols.

Given a finite word problem $\bigwedge_{i\in I} s_i=t_i \vdash s=t$
where the function symbols are either uninterpreted or AC,
AC-completion attempts to transform the conjunction $\bigwedge_{i\in
  I} s_i=t_i$ into a finitely terminating, confluent term rewriting
system $R$ whose reductions preserve identity. The rewriting system
$R$ serves as a decision procedure for validating $s=t$ modulo AC: the 
equation holds if and only if the normal forms of $s$ and $t$ w.r.t
$R$ are equal modulo AC. Furthermore, when its input contains only
ground equations, AC-completion terminates and outputs a convergent
rewriting system~\cite{marche91rta}.

Unfortunately, AC reasoning is only a part of the automated deduction
problem, and what we really need is to decide formulas combining AC
symbols and other theories.  For instance, in practice, we are
interested in deciding finite ground word problems which contain a
mixture of uninterpreted, interpreted and AC function symbols, as in
the following assertion
\[
\begin{array}[h]{lc}
  u(a,c_2-c_1) =  a ~\wedge~ u(e_1,e_2) - f(b) =
  u(d,d) ~\wedge~& \\ [0.6em]
  d= c_1+1 ~\wedge~ e_2= b ~\wedge~ u(b,e_1) =
  f(e_2) ~\wedge~ c_2 =  2*c_1 + 1  
\end{array}
 \vdash
  a =  u(a,0),
\]
where $u$ is an AC symbol, $+$, $-$, $*$ and the numerals are from the
theory of linear arithmetic, $f$ is an uninterpreted function symbol
and the other symbols are uninterpreted constants.  A combination of
AC reasoning with linear arithmetic and the free theory $\CE$ of
equality is necessary to prove this formula.  Linear arithmetic is
used to show that $c_2-c_1 = c_1+1$ so that $(i)$ $u(a,c_1+1)=a$
follows by congruence. Independently, $e_2=b$ and $d=c_1+1$ imply
$(ii)$ $u(c_1+1,c_1+1) = 0$ by congruence, linear arithmetic and
commutativity of $u$. AC reasoning can finally be used to conclude
that $(i)$ and $(ii)$ imply that $u(a,c_1+1,c_1+1)$ is equal to both
$a$ and $u(a,0)$.

There are two main methods for combining decision procedures for
disjoint theories.  First, the Nelson-Oppen
approach~\cite{nelson79pls} is based on a variable abstraction
mechanism and the exchange of equalities between shared
variables. Second, the Shostak's algorithm~\cite{shostak84jacm}
extends a congruence closure procedure with theories equipped with
canonizers and solvers, \emph{i.e.}  procedures that compute
canonical forms of terms and solve equations, respectively.
While ground AC-completion can be easily combined with other decision
procedures by the Nelson-Oppen method, it cannot be directly
integrated in the Shostak's framework since it actually does not
provide a solver for the AC theory.

In this paper, we investigate the integration of Shostak theories in
ground AC-completion. We first introduce a new notion of rewriting
called \emph{canonized} rewriting which adapts normalized rewriting to
cope with canonization. Then, we present a modular extension of ground
AC-completion for deciding formulas in the combination of the theory
of equality with user-defined AC symbols, uninterpreted symbols and an
arbitrary signature disjoint Shostak theory \X. The main ideas of our
integration are to substitute standard rewriting by canonized
rewriting, using a global canonizer for AC and \X, and to replace the
equation orientation mechanism found in ground AC-completion with the
solver for \textsf{X}.

AC-completion has been studied for a long time
in the rewriting community~\cite{lankford77a,peterson81}. A generic
framework for combining completion with a generic built-in equational
theory $E$ has been proposed in~\cite{jouannaud86siam}. Normalized
completion~\cite{marche96jsc} is designed to use a modified rewriting
relation when the theory $E$ is equivalent to the union of the AC
theory and a convergent rewriting system $\mathcal{S}$.  In this
setting, rewriting steps are only performed on
$\mathcal{S}$-normalized terms. \ACX{} can be seen as an
adaptation of ground normalized completion to efficiently handle the
theory $E$ when it is equivalent to the union of the AC theory and a
Shostak theory {\X}. In particular, $\mathcal{S}$-normalization is
replaced by the application of the canonizer of {\X}. This modular
integration of {\X} allows us to reuse proof techniques of ground 
AC-completion~\cite{marche91rta} to show the correctness
of \ACX{}. 

Tiwari~\cite{tiwari09frocos} efficiently combined equality and AC
reasoning in the Nelson-Oppen framework. Kapur~\cite{kapur97rta2} used
ground completion to demystify Shostak's congruence closure algorithm
and Bachmair \textit{et al.}~\cite{bachmair03jar} compared its
strategy with other ones into an abstract congruence closure
framework. While the latter approach can also handle AC symbols, none
of these works formalized the integration of Shostak theories into
(AC) ground completion.

\textit{Outline.} Section~{\ref{sec:prelim}} recalls standard ground AC
completion. Section~\ref{sec:shostak} is devoted to Shostak theories
and global canonization. Section~\ref{sec:acx} presents the
\ACX{} algorithm and illustrates its use through an example. The
correctness of \ACX{} is detailed in
Section~\ref{sec:correctness:annexe}. In
Section~\ref{sec:termabstraction}, we show that a simple preprocessing
step allows us to use a partial multiset ordering instead of a full
AC-compatible reduction ordering. Experimental results are presented
in Section~\ref{lab:experiments}. Using a simple example, we illustrate in
Section~\ref{sec:limit} how the instantiation mechanism of \ergo{}
has to be extended modulo AC in order to fully integrate \ACX{}
as a core decision procedure for our SMT solver. Conclusion and
future works are presented in Section~\ref{sec:conclusion}.

\section{Ground AC-Completion}
\label{sec:prelim}
In this section, we first briefly recall the usual notations and
definitions of \cite{baader98book,dershowitz88} for term rewriting
modulo AC.  Then, we give the usual set of inference rules for ground
AC-completion procedure and we illustrate its use through an example.

Terms are built from a signature $\Sigma = \Sigma_{AC} \uplus
\Sigma_{\CE}$ of AC and uninterpreted symbols, and a set of variables
$\CX$ yielding the term algebra $\TFX$. The range of letters $a\ldots
f$ denotes uninterpreted symbols, $u$ denotes an AC function symbol,
$s$, $t$, $l$, $r$ denote terms, and $x$, $y$, $z$ denote
variables. Viewing terms as trees, subterms within a term $s$ are
identified by their positions. Given a position $p$, $s|_p$ denotes
the subterm of $s$ at position $p$, and $s[r]_p$ the term obtained by
replacement of $s|_p$ by the term $r$. We will also use the notation
$s(p)$ to denote the symbol at position $p$ in the tree, and the root
position is denoted by $\Lambda$. Given a subset $\Sigma'$ of
$\Sigma$, a subterm $t|_p$ of $t$ is a $\Sigma'$-alien of $t$ if
$t(p)\not\in \Sigma'$ and $p$ is minimal \emph{w.r.t} the prefix word
ordering\footnote{Notice that according to this definition, a variable
  may be a $\Sigma'$-alien.}. We write $\CA_{\Sigma'}(t)$ the multiset
of $\Sigma'$-aliens of $t$.

A substitution is a partial mapping from variables to
terms. Substitutions are extended to a total mapping from terms to
terms in the usual way. We write $t\sigma$ for the application of a
substitution $\sigma$ to a term $t$. A well-founded
quasi-ordering~\cite{dershowitz82tcs} on terms is a reduction
quasi-ordering if $s \preceq t$ implies $s\sigma \preceq t\sigma$ and
$l[s]_p \preceq l[t]_p$, for any substitution $\sigma$, term $l$ and
position $p$. A quasi-ordering $\preceq$ defines an equivalence
relation $\simeq$ as $\preceq\cap \succeq$ and a partial ordering
$\prec$ as $\preceq \cap \not\succeq$.

An equation is an unordered pair of terms, written $s \approx t$. The
variables contained in an equation, if any, are understood as being
universally quantified. Given a set of equations $E$, the equational
theory of $E$, written $=_E$, is the set of equations that can be
obtained by reflexivity, symmetry, transitivity, congruence and
instances of equations in $E$\footnote{The equational theory of the
free theory of equality $\CE$, defined by the empty set of equations,
is simply denoted $=$.}. The word problem for $E$ consists in
determining if, given two ground terms $s$ and $t$, the equation
$s \approx t$ is in $=_E$, denoted by $s =_E t$. The word problem for
$E$ is ground when $E$ contains only ground equations. An equational
theory $=_E$ is said to be \emph{inconsistent} when $s =_E t$,
for \emph{any} $s$ and $t$.

A rewriting rule is an oriented equation, usually denoted by $l
\rightarrow r$. A term $s$ rewrites to a term $t$ at position $p$ by
the rule $l\rightarrow r$, denoted by $s\rightarrow_{l\rightarrow r}^p
t$, iff there exists a substitution $\sigma$ such that $s|_p =
l\sigma$ and $t=s[r\sigma]_p$. A rewriting system $R$ is a set of
rules. We write $s\rightarrow_R t$ whenever there exists a rule
$l\rightarrow r$ of $R$ such that $s$ rewrites to $t$ by $l\rightarrow
r$ at some position. A normal form of a term $s$ w.r.t to $R$ is a
term $t$ such that $s \rightarrow_R^* t$ and $t$ cannot be rewritten
by $R$. The system $R$ is said to be {\em convergent} whenever any
term $s$ has a unique normal form, denoted $\nf{R}{s}$, and does not
admit any infinite reduction. Completion~\cite{knuth70} aims at
converting a set $E$ of equations into a convergent rewriting system
$R$ such that the sets $=_E$ and $\{s \approx t ~|~ \nf{R}{s} =
\nf{R}{t}\}$ coincide. Given a suitable reduction ordering on terms,
it has been proved that completion terminates when $E$ is
ground~\cite{lankford75}.

\paragraph{Rewriting modulo AC} Let $=_{AC}$ be the equational theory
obtained from the set:
\[
   AC = \bigcup_{u\in\Sigma_{AC}}\{\, u(x,y) \approx u(y,x),\,
   u(x,u(y,z)) \approx u(u(x,y),z) \,\}.
\]
In general, given a set $E$ of equations, it has been shown that no
suitable reduction ordering allows completion to produce a convergent
rewriting system for $E \cup AC$. When $E$ is ground, an alternative
consists in in-lining AC reasoning both in the notion of rewriting
step and in the completion procedure. 

Rewriting modulo AC is directly related to the notion of matching
modulo AC as shown by the following example. Given a rule
$u(a,u(b,c)))\rightarrow t$, we would like the following reductions to
be possible:
\begin{enumerate}[(1)]
\item $f(u(c,u(b,a)),d) \rightarrow f(t,d)$,
\item $u(a,u(c,u(d,b))) \rightarrow u(t,d)$.
\end{enumerate}
Associativity and commutativity of $u$ are needed in $(1)$ for the
subterm $u(c,u(b,a))$ to match the term $u(a,u(b,c))$, and in $(2)$
for the term $u(a,u(c,u(d,b)))$ to be seen as $u(u(a,u(b,c)),d)$, so
that the rule can be applied. More formally, this leads to the
following definition.
\begin{defi}[Ground rewriting modulo AC]
  \label{def:rewriteac}
  A term $s$ rewrites to a term $t$ modulo AC at position $p$ by the
  rule \mbox{$l\rightarrow r$}, denoted by $s\rightarrow_{AC\backslash
    l\rightarrow r}^p t$, iff one of the following conditions holds:
  \begin{enumerate}[(1)]
    \item $s|_p =_{AC} l$ and $t=s[r]_p$, 
    \item $l(\Lambda) = u$ and there exists a term
  $s'$ such that $s|_p =_{AC} u(l,s')$ and $t=s[u(r,s')]_p$.
\end{enumerate}
\end{defi}

In order to produce a convergent rewriting system, ground
AC-completion requires a well-founded reduction quasi-ordering
$\preceq$ total on ground terms with an underlying equivalence
relation which coincides with $=_{AC}$. Such an ordering will be
called a total ground AC-reduction ordering.

The inference rules for ground AC-completion are given in
Figure~\ref{fig:rule:ac}. The rules describe the evolution of the
state of a procedure, represented as a configuration
$\complbis{E}{R}$, where $E$ is a set of ground equations and $R$ a
ground set of rewriting rules. The initial state is
$\complbis{E_0}{\emptyset}$ where $E_0$ is a given set of ground 
equations. \trivial{} removes an equation $u \approx v$ from $E$ when
$u$ and $v$ are equal modulo AC. \orient{} turns an equation into a
rewriting rule according to a given total ground AC-reduction ordering
$\preceq$. $R$ is used to rewrite either side of an equation
(\simplify), and to reduce right hand side of rewriting rules
(\compose{}).  Given a rule $l \rightarrow r$, \collapse{} either
reduces $l$ at an inner position, or replaces $l$ by a term smaller
than $r$. In both cases, the reduction of $l$ to $l'$ may influence
the orientation of the rule $l' \rightarrow r$ which is added to $E$
as an equation in order to be re-oriented. Finally, \deduce{} adds
equational consequences of rewriting rules to $E$. For instance, if
$R$ contains two rules of the form $u(a,b) \rightarrow s$ and $u(a,c)
\rightarrow t$, then the term $u(a,u(b,c))$ can either be reduced to
$u(s,c)$ or to the term $u(t,b)$. The equation $u(s,c) \approx
u(t,b)$, called \emph{critical pair}, is thus necessary for ensuring
convergence of $R$. Critical pairs of a set of rules are computed by
the following function ($a^\mu$ stands for the maximal term
w.r.t. size enjoying the assertion):
\[
\fheadcp(R) = \left\{~
  u(b,r') \approx u(b',r) ~\left|~
 \begin{array}{l}
   l \rightarrow r\in R,~~ l' \rightarrow r' \in R ~\\
   \exists\, a^\mu:~l =_{_{AC}} u(a^\mu,b) ~\land~ l' =_{_{AC}} u(a^\mu,b')
 \end{array}
\right.
\right\}.
\]
\begin{figure}[htbp]
  \fbox{
  \begin{minipage}{\textwidth}
    \[
    \hspace*{-3cm}
    \begin{array}[h]{c}
      \inferrule*[left={\trivial},
      Right= \hbox{$s\;=_{AC}\, t$}]
      {\compl{E\cup \{\,s \approx t\,\}}{R}
      }{
      \compl{E}{R}}\\[2mm]
    \inferrule*[left={\orient},
      Right= \hbox{$t \prec s$}]
      {\compl{E\cup \{\,s \approx t\,\}}{R}
      }{
        \compl{E}{R \cup \{\,s\rightarrow t\,\}}}\\[2mm]
      \inferrule*[left={\simplify},
      Right= \hbox{$s \rightarrow_{AC\backslash R} s'$}]
      {\compl{E\cup \{\,s \approx t\,\}}{R}
      }{
      \compl{E\cup \{\,s' \approx t\,\}}{R}}\\[2mm]
      \inferrule*[left=\compose, 
      Right=$r \rightarrow_{AC\backslash R} r'$]{\compl{E}{R\cup\{\,l\rightarrow
          r\,\}}
      }{
        \compl{E}{R\cup\{\,l\rightarrow r'\,\}}}\\[2mm]
      \inferrule*[left=\collapse, 
      Right=\mbox{
        $\left\{\begin{array}[h]{l} 
          l \rightarrow_{AC\backslash g\rightarrow d} l'\\[1mm]
          g \prec l \;\vee \;  (g\simeq l \wedge d\prec r)
        \end{array}\right.$}]{\compl{E}
      {R\cup\{\,g\rightarrow d,\, l\rightarrow r\,\}}
      }{
        \compl{E\cup\{\,l'\approx r\,\}}{R\cup\{\,g\rightarrow d\,\}}}\\[2mm]
      \inferrule*[left=\deduce, 
      Right=\mbox{$s\approx t\in\fheadcp(R)$}]
      {\compl{E}{R}
      }{
        \compl{E\cup\{\,s\approx t\,\}}{R}}
    \end{array}\]
    \end{minipage}}
  \caption{Inference rules for ground AC-completion.}
  \label{fig:rule:ac}
\end{figure}
\paragraph{Example.}
To get a flavor of ground AC-completion, consider a modified version
of the assertion given in the introduction, where the arithmetic part
has been removed (and uninterpreted constant symbols renamed for the
sake of simplicity)
\[
u(a_1,a_4) \approx a_1, u(a_3,a_6) \approx u(a_5,a_5), a_5\approx a_4, a_6\approx a_2 
\vdash a_1 \approx u(a_1,u(a_6,a_3)).
\]
The precedence $a_1 \prec_p \cdots \prec_p a_6 \prec_p u$ defines an
AC-RPO ordering on terms~\cite{nieuwenhuis93rta} which is suitable for
ground AC-completion. The table in Figure~\ref{fig:ex1} shows the
application steps of the rules given in Figure~\ref{fig:rule:ac} from
an initial configuration 
\[
\langle\, \{ u(a_1,a_4) \approx a_1,
u(a_3,a_6) \approx u(a_5,a_5), a_5\approx a_4, a_6\approx
a_2 \}~|~\emptyset\,\rangle
\] 
to a final configuration \mbox{$\langle\,\emptyset~|~R_f\,\rangle$}, where $R_f$
is the set of rewriting rules $\{1, 3, 5, 7, 10\}$. It can be checked
that $\nf{R_f}{a_1}$ and 
$\nf{R_f}{u(a_1,u(a_6,a_3))}$ are identical.

\begin{figure}[h]
  \begin{center}
    \begin{minipage}[h]{12cm}
      \renewcommand{\arraystretch}{1.4}
      \renewcommand{\tabcolsep}{3mm}
      {
        \begin{tabular}[h]{|p{0.2cm}|p{5cm}|p{5cm}|}
          \hline
          1 & $\mathbf{u(a_1,a_4) \rightarrow a_1}$ & 
          \orientbis{} $u(a_1,a_4)\approx a_1$\\
          \hline
          2 & $u(a_3,a_6) \rightarrow u(a_5,a_5)$ & 
          \orientbis{} $u(a_3,a_6) \approx u(a_5,a_5)$ \\
          \hline
          3 & $\mathbf{a_5 \rightarrow a_4}$ & 
          \orientbis{} $a_5 \approx a_4$ \\
          \hline
          4 & $u(a_3,a_6) \rightarrow u(a_4,a_4)$ & 
          \composebis{} $2$ and $3$ \\
          \hline
          5 & $\mathbf{a_6 \rightarrow a_2}$ & 
          \orientbis{} $a_6 \approx a_2$ \\
          \hline
          6 & $u(a_3,a_2) \approx u(a_4,a_4)$ & 
          \collapsebis{} $4$ and $5$\\
          \hline
          7 & $\mathbf{u(a_4,a_4) \rightarrow u(a_3,a_2)}$ & 
          \orientbis{} $6$ \\
          \hline
          8 & $u(a_1,a_4) \approx u(a_1,u(a_3,a_2))$ & 
          \deducebis{} from $1$ and $7$ \\
          \hline
          9 & $a_1 \approx u(a_1,u(a_3,a_2))$ & 
          \simplifybis{} 8 by 1\\
          \hline
          10 & $\mathbf{u(a_1,u(a_3,a_2)) \rightarrow a_1}$ & 
          \orientbis{} $9$\\
          \hline
        \end{tabular}
      }
      \caption{Ground AC-completion example.}
      \label{fig:ex1}
    \end{minipage}
  \end{center}
\end{figure}

\section{Shostak Theories and Global Canonization}
\label{sec:shostak}
In this section, we recall the notions of canonizers and solvers
underlying Shostak theories and show how to obtain a global canonizer
for the combination of the theories $\CE$ and AC with an 
arbitrary signature disjoint Shostak theory \X.

From now on, we assume given a theory {\X} with a signature
$\Sigma_{\X}$. A canonizer for {\X} is a function $\canX$ that
computes a unique normal form for every term such that $s =_\X
t \ \mbox{iff} \ \canX(s)=\canX(t)$. A solver for {\X} is a function
$\fsolveX$ that solves equations between $\Sigma_{\X}$-terms. Given an
equation $s\approx t$, $\fsolveX(s\approx t)$ either returns a special
value $\bot$ when $s\approx t\cup \X$ is inconsistent, or an
equivalent substitution. A Shostak theory {\X} is a theory with a
canonizer and a solver which fulfill some standard properties given
for instance in~\cite{krstic-conchon-05}.

Our combination technique is based on the integration of a Shostak
theory {\X} in ground AC-completion. 
From now on, we assume that terms are built from a signature $\Sigma$
defined as the union of the disjoint signatures $\Sigma_{AC}$,
$\Sigma_{\CE}$ and $\Sigma_\X$. We also assume a total ground
AC-reduction ordering $\preceq$ defined on $\TFX$ used later on for
completion.
The combination mechanism requires defining both a global canonizer
for the union of $\CE$, AC and {\X}, and a wrapper of $\fsolveX$ to
handle heterogeneous equations.
These definitions make use of a global one-to-one mapping $\alpha
: \TF\rightarrow\CX$ (and its inverse mapping $\rho$) and are based on
a variable abstraction mechanism which computes the \emph{pure}
$\Sigma_{\X}$-part $\make{t}$ of a heterogeneous term $t$ as follows:
\[
\make{t} =
\left\{
  \begin{array}[h]{lr}
    f(\make{\vec s}) &\qquad \mbox{when}~t = f(\vec s)
    ~\mbox{and}~ f\in\Sigma_{\X},\\[1em]
    \alpha(t) &\mbox{otherwise}.
\end{array}
\right.
\]
The canonizer for AC defined in \cite{hullot79} is based on flattening
and sorting techniques which simulate associativity and commutativity,
respectively. For instance, the term $u(u(u'(c,b),b),c)$ is first
flattened to $u(u'(c,b),b,c)$ and then sorted\footnote{For instance,
using the AC-RPO ordering based on the precedence $b\prec_p c \prec_p
u'$.} to get the term $u(b,c,u'(c,b))$. It has been formally proved
that this canonizer solves the word problem for
AC~\cite{contejean04rta}. However, this definition implies a
modification of the signature $\Sigma_{AC}$ where arity of AC symbols
becomes variadic. Using such canonizer would impact the definition of
AC-rewriting given in Section~\ref{sec:prelim}. In order to avoid such
modification we shall define an equivalent canonizer that builds
degenerate trees instead of flattened terms. For instance, we would
expect the normal form of $u(u(u'(c,b),b),c)$ to be
$u(b,u(c,u'(c,b)))$. Given a signature $\Sigma$ which contains
$\Sigma_{AC}$ and any total ordering $\unlhd$ on terms, we define
$\canAC$ by: 
\[
\begin{array}{lcl}
 \canAC(x) &\,=\,& x \hfill \mathrm{when}~~ x\in \CX,\\[0.2em]
 \canAC(f(\vec v)) & \,=\, & 
 f(\canAC(\vec v)) 
 \hfill \mathrm{when}~~ f\not\in \Sigma_{AC},\\[0.2em]
 \canAC(u(t_1,t_2)) &\,=\,& u(s_1,u(s_2,\ldots,u(s_{n-1},s_n)\ldots))\\[0.2em]
 ~ & ~ & \mathrm{where}~ t_i' = \canAC(t_i) ~\mathrm{for}~ i \in
 [1,2]\\[0.2em]  
 ~ & ~ & \mathrm{and}~ \multiset{s_1,\ldots,s_n}
 = \CA_{\{u\}}(t_1') \cup \CA_{\{u\}}(t_2')\\[0.2em] 
 ~ & ~ & \mathrm{and}~ s_{i} \unlhd s_{i+1} ~\mathrm{for}~i\in [1,n-1],
\hfill  ~~\mathrm{when}~~ u\in\Sigma_{\AC}. 
\end{array}
\]
We can easily show that $\canAC$ enjoys the standard properties
required for a canonizer. The proof that $\canAC$ solves the word
problem for AC follows directly from the one given
in~\cite{contejean04rta}. 

Using the technique described in~\cite{krstic-conchon-05}, we define
our global canonizer $\can$ which combines $\canX$ with $\canAC$ as
follows:
\[
 \begin{array}{lcll}
 \can(x) & \,=\, & x & \quad\mathrm{when}~~ x\in \CX,\\[0.2em]
 \can(f(\vec v)) & \,=\, & f(\can (\vec v)) &
 \quad\mathrm{when}~f\in\Sigma_{\CE},\\[0.2em]
 \can(u(s,t)) & \,=\, & \canAC(u(\can (s),\can (t))) &
\quad\mathrm{when}~~ u\in\Sigma_\AC,\\[0.2em]
 \can(f_{_\X}(\vec v)) & \,=\, & \canX(f_{_\X}(\make{\can(\vec
 v)}))\rho &  \quad\mathrm{when}~~ 
 f_{_\X}\in\Sigma_\X.\\
 \end{array}
 \]
Again, the proofs that $\can$ solves the word problem for the union
$\CE$, AC and {\X} and enjoys the standard properties required for a
canonizer  are similar to those given in~\cite{krstic-conchon-05}. The
only difference is that $\canAC$ directly works on the signature
$\Sigma$, which avoids the use of a variable abstraction step when
canonizing a mixed term of the form $u(t_1,t_2)$ such that $u \in
\Sigma_{AC}$.

Using the same mappings $\alpha$, $\rho$ and the abstraction function,
the wrapper $\fsolve$ can be easily defined by:
\[
\fsolve(s\approx t) = \left\{
\begin{array}{lcl}
  \bot & \quad &
  \mbox{if~~} \fsolveX(\make{s}\approx \make{t}) = \bot,
  \\[0.1em] 
  \{~x_i\rho \to t_i\rho~\} & \quad &
  \mbox{if~~} \fsolveX(\make{s}\approx \make{t}) = 
  \{x_i \!\approx\!  t_i\}.
\end{array}
\right.
\]
In order to ensure termination of \ACX{}, the global canonizer
and the wrapper must be compatible with the ordering $\preceq$ used by
AC-completion, that is: 
\begin{lem}\hfill
\begin{enumerate}[\em(1)]
\item $\forall t\in\TF,\  \can(t) \preceq t$, 
\item $\forall s,t\in\TF,\ ~\mbox{if}~\fsolve(s\approx t)
= \bigcup\{p_i \rightarrow v_i\} ~\mbox{then}~ v_i \prec p_i$.
\end{enumerate}
\end{lem}
We can prove that the above properties hold when the theory \X{} enjoys
the following local compatibility properties:
\begin{axiom}\label{ax:solve}\hfill
\begin{enumerate}[\em(1)]
\item $\forall t\in\TF,\ \canX(\make{t}) \preceq \make{t}$,
\item $\forall s,t\in\TF, \ \mbox{if}~\fsolveX(\make{s}\approx \make{t})
= \bigcup\{x_i \approx t_i\} ~\mbox{then}~ t_{i}\rho \prec x_i\rho$.
\end{enumerate}
\end{axiom}
To fulfill this axiom, AC-reduction ordering can be chosen as an AC-RPO
ordering~\cite{nieuwenhuis93rta} based on a precedence relation
$\prec_p$ such that $\Sigma_{\X} \prec_p \Sigma_{\CE} \cup
\Sigma_{AC}$. From now on, we assume that $\X$ is locally compatible
with $\preceq$.

\paragraph{Example.} To solve the equation $u(a,b) + a \approx 0$, we
use the abstraction 
\[
\alpha = \{u(a,b) \mapsto x,~ a\mapsto y\}
\]
and call $\fsolveX$ on $x + y \approx 0$. Since $a \prec u(a,b)$, the only
solution which fulfills the axiom above is $\{x \approx -y\}$. We
apply $\rho$ and get the set $\{u(a,b) \rightarrow -a\}$ of rewriting
rules.

\section{Ground AC-Completion Modulo \texorpdfstring{$\mathsf{X}$}~}
\label{sec:acx}

In this section, we present the \ACX{} algorithm which extends
the ground AC-completion procedure given in
Section~\ref{sec:prelim}. For that purpose, we first adapt the notion
of ground AC-rewriting to cope with canonizers. Then, we show how to
refine the inference rules given in Figure~\ref{fig:rule:ac} to reason
modulo the equational theory induced by a set $E$ of ground equations
and the theories $\CE$, AC and \X.

\subsection{Canonized Rewriting}
From the rewriting point of view, a canonizer behaves like a
convergent rewriting system: it gives an effective way of computing
normal forms. Thus, a natural way for integrating $\can$ in ground
AC-completion is to extend normalized rewriting~\cite{marche96jsc}.
\begin{defi}
\label{def:srs}
Let $\can{}$ be a canonizer. A term $s$ $\can$-rewrites to a term $t$
at position $p$ by the rule \mbox{$l\rightarrow r$}, denoted by
$s\rightsquigarrow_{l\rightarrow  r}^p t$, iff
\[
     s\rightarrow_{AC\backslash l\rightarrow r}^p t'
     \qquad\mathrm{and}\qquad 
     \can(t') = t.
\]
\end{defi}

\paragraph{Example.}
Using the usual canonizer $\can_\CA$ for linear arithmetic and the
rule $\gamma: u(a,b)\to a$, the term $f(a + 2 * u(b,a))$ 
$\can_\CA$-rewrites to $f(3*a)$ by $\rightsquigarrow_{\gamma}$ as
follows:
\[
f(a + 2 * u(b,a)) \rightarrow_{AC\backslash \gamma}  f(a + 2 * a)
~~\mbox{and}~~
\can_\CA(f(a + 2 * a)) = f(3 * a).
\]
\begin{lem}
\label{lem:rsa_sound}
$\forall ~s,~t. ~~~ s \rightsquigarrow_{l\rightarrow r}
t \implies s =_{AC,\X,l \approx r} t. $ \qed
\end{lem}

\subsection{The \texorpdfstring{\textsf{AC($\mathsf{X}$)}} ~~Algorithm}
The first step of our combination technique consists in replacing the
rewriting relation found in completion by canonized rewriting. This
leads to the rules of \ACX{} given in Figure~\ref{fig:rule:acx}.
The state of the procedure is a pair $\complbis{E}{R}$ of equations
and rewriting rules. The initial configuration is 
$\complbis{E_0}{\emptyset}$ where $E_0$ is supposed to be a set of
equations between canonized terms. Since \ACX{}'s rules only
involve canonized rewriting, the algorithm maintains the invariant
that terms occurring in $E$ and $R$ are in canonical forms. \trivial{}
thus removes an equation $u \approx v$ from $E$ when $u$ and $v$ are
syntactically equal. A new rule {\bottom} is used to detect
inconsistent equations. Similarly to normalized completion,
integrating the global canonizer $\can$ in rewriting is not enough to 
fully extend ground AC-completion with the theory {\X}: in both cases
the orientation mechanism has to be adapted . Therefore, the second
step consists in integrating the wrapper {\fsolve} in the {\orient}
rule. The other rules are much similar to those of ground
AC-completion except that they use the relation $\rsa_R$ instead of
$\rightarrow_{AC\backslash R}$.
\begin{figure}[htbp]
  \fbox{
  \begin{minipage}{0.97\textwidth}
    \[
    \hspace*{-2.5cm}
    \begin{array}[h]{c}
      \inferrule*[left={\trivial{}},
      Right= \hbox{$s\;=\, t$}]
      {\complbis{E\cup \{\,s \approx t\,\}}{R}
      }{
      \complbis{E}{R}}
      \qquad\qquad
      \inferrule*[left=\bottom{},
      Right=\mbox{$\fsolve(s,t) = \bot$}]
      {\complbis{E\cup\{\,s\approx t\,\}}{R}
      }{
        \bot}
      \\[4mm]
      \inferrule*[left=\orient{},
      Right=\mbox{$\fsolve(s,t) \neq \bot$}]
      {\complbis{E\cup\{\,s\approx t\,\}}{R}
      }{
        \complbis{E}{R\cup \fsolve(s,t)}}
      \\[4mm]
      \quad\qquad\inferrule*[left={\simplify{}},
      Right= \hbox{$s \rightsquigarrow_{R} s'$}]
      {\complbis{E\cup \{\,s \approx t\,\}}{R}
      }{
      \complbis{E\cup \{\,s' \approx t\,\}}{R}}
      \qquad\qquad\qquad~
      \hspace*{-0.5cm}
      \inferrule*[left=\compose{}, 
      Right=$r \rightsquigarrow_{R} r'$]{\complbis{E}{R\cup\{\,l\rightarrow
          r\,\}}
      }{
        \complbis{E}{R\cup\{\,l\rightarrow r'\,\}}}
      \\[4mm]
      \inferrule*[left=\collapse{}, 
      Right=\mbox{
        $\left\{\begin{array}[h]{l} 
          l \rightsquigarrow_{g\rightarrow d} l'\\[1mm]
          g  \prec l \;\vee \;  (g\simeq l \wedge d \prec r)
        \end{array}\right.$}]{\complbis{E}
      {R\cup\{\,g\rightarrow d,\, l\rightarrow r\,\}}
      }{
        \complbis{E\cup\{\,l'\approx r\,\}}{R\cup\{\,g\rightarrow
          d\,\}}}
      \\[4mm]
      \inferrule*[left=\deduce{}, 
      Right=\mbox{$s\approx t\in\fheadcp(R)$}]
      {\complbis{E}{R}
      }{
        \complbis{E\cup\{\,s\approx t\,\}}{R}}
    \end{array}\]
    \end{minipage}}
  \caption{Inference rules for ground AC-completion modulo \X.}
  \label{fig:rule:acx}
\end{figure}
\paragraph{Example.}
We illustrate \ACX{} on the example given in the introduction:
\[
\begin{array}[h]{lc}
  u(a,c_2-c_1) \approx  a \,\wedge\, u(e_1,e_2) - f(b) \approx
  u(d,d) \,\wedge& \\ 
  d\approx c_1+1 \,\wedge\, e_2\approx b \,\wedge\, u(b,e_1) \approx
  f(e_2) \,\wedge\, c_2 \approx  2*c_1 + 1  
\end{array}
 \vdash
  a \approx  u(a,0).
\]
The table given in Figure~\ref{fig:ex2} shows the application of the
rules of \textsf{AC(\X)} on the example when {\X} is instantiated by
linear arithmetic. We use an AC-RPO ordering based on the precedence
$1 \prec_p 2 \prec_p a \prec_p b \prec_p c_1 \prec_p c_2 \prec_p
d \prec_p e_1 \prec_p e_2 \prec_p f \prec_p u$. The procedure
terminates and produces a convergent rewriting system $R_f = \{3,5,9,10,
11,13,16\}$. Using $R_f$, we can check that $a$ and $u(a,0)$
$\can$-rewrite to the same normal form. 

\begin{figure}[h]
  \begin{center}
    \begin{minipage}[h]{14cm}
      \renewcommand{\arraystretch}{1.4}
      \renewcommand{\tabcolsep}{3mm}
      {
        \begin{tabular}[h]{|p{0.2cm}|p{6cm}|p{6cm}|}
      \hline
      1 & $u(a,c_2-c_1) \to a$ & 
      \xsolvebis{} $u(a,c_2-c_1) \approx  a$\\
      \hline
      2 & $u(e_1,e_2) \to u(d,d) + f(b)$ & 
      \xsolvebis{} $u(e_1,e_2) - f(b) \approx  u(d,d)$\\
      \hline
      3 & $\mathbf{d\to c_1+1}$ & 
      \xsolvebis{} $d\approx c_1+1$\\
      \hline
      4 & $u(e_1,e_2) \to u(c_1+1,c_1+1) + f(b)$ & 
      \composebis{} $2$ and $3$\\
      \hline
      5 & $\mathbf{e_2 \to b}$ & 
      \xsolvebis{} $e_2\approx b$\\
      \hline
      6 & $u(b,e_1) \approx u(c_1+1,c_1+1) + f(b)$ & 
      \collapsebis{} $4$ and $5$\\
      \hline
      7 & $u(b,e_1) \to u(c_1+1,c_1+1) + f(b)$ & 
      \xsolvebis{} $u(b,e_1) \approx u(c_1+1,c_1+1) + f(b)$\\
      \hline
      8 & $u(c_1+1,c_1+1) + f(b) \approx  f(b)$ & 
      \simplifybis{} $u(b,e_1) \approx  f(e_2)$ by $5$ and $7$\\
      \hline
      9 & $\mathbf{u(c_1+1,c_1+1) \to 0}$ & 
      \xsolvebis{} $u(c_1+1,c_1+1) + f(b) \approx  f(b)$\\
      \hline
      10 & $\mathbf{u(b,e_1) \to f(b)}$ & 
      \composebis{} $7$ and $9$\\
      \hline
      11 & $\mathbf{c_2 \to  2*c_1 + 1}$ & 
      \xsolvebis{} $c_2 \approx  2*c_1 + 1$\\
      \hline
      12 & $u(a,c_1+1) \approx a$ & 
      \collapsebis{} $1$ and $11$\\
      \hline
      13 & $\mathbf{u(a,c_1+1) \to a}$ & 
      \xsolvebis{} $u(a,c_1+1) \approx a$\\
      \hline
      14 & $u(0,a) \approx u(a,c_1+1)$ & 
      \deducebis{} from $9$ and $13$\\
      \hline
      15 & $u(0,a) \approx a$ & 
      \simplifybis{} $14$ by $13$\\
      \hline
      16 & $\mathbf{u(0,a) \to a}$ & 
      \xsolvebis{} $15$\\
      \hline
    \end{tabular}
    \caption{\ACX{} on the running example.}  
  \label{fig:ex2}
  }
\end{minipage}
\end{center}
\end{figure}

\section{Correctness}
\label{sec:correctness:annexe}

In this section, we give detailed proofs for the correctness of
\ACX{}. This property is stated by the theorem below and its
proof is based on three intermediate theorems, stating respectively
soundness, completeness and termination.  

As usual, in order to enforce correctness, we cannot use any (unfair)
strategy. We say that a strategy is {\em strongly fair} when no
possible application of an inference rule is infinitely delayed 
and \orient{} is only applied over fully
reduced terms. 

\begin{thm}
  Given a set $E$ of ground equations, the application of the rules of
  \ACX{} under a strongly fair strategy terminates and either
  produces $\bot$ when $E\cup AC \cup \X$ is inconsistent, or yields a
  final configuration $\complbis{\emptyset}{R}$ such that: 
\[
\forall
  s,t\in\TF.\ s =_{_{E,AC,X}} t \iff \can{(s)}\downsquigarrow_{_R} =
  \can{(t)}\downsquigarrow_{_R}.
\]
\end{thm}
In the following, we shall consider a fixed run of the
completion procedure 
\[
\complbis{E_0}{\emptyset} \rightarrow \complbis{E_1}{R_1} \rightarrow
\ldots \rightarrow \complbis{E_n}{R_n}\rightarrow \complbis{E_{n+1}}{R_{n+1}}
\rightarrow \ldots
\]
starting from the initial configuration
$\complbis{E_0}{\emptyset}$. We denote $R_\infty$ (resp. $E_\infty$)
the set of all encountered rules $\bigcup_{n}R_n$ (resp. equations
$\bigcup_{n}E_n$) and $\R_\omega$ (resp. $E_\omega$) the set of
persistent rules $\bigcup_{n}\bigcap_{i \geq n}R_i$ (resp. equations
$\bigcup_{n}\bigcap_{i \geq n}E_i$). 

The strongly fair strategy requirement implies in
particular that $\fheadcp{(R_\omega)}\subseteq E_\infty$, $E_\omega =
\emptyset$ and $R_\omega$ is inter-reduced, that is none of its rules
can be collapsed or composed by another one.
Due to the assumptions made over
$\can_\X$ and $\prec$, the following valid properties will be
continuously used in the proofs:
\[
\begin{array}{l}
\forall t.\ \can(t)\preceq t,\\
\forall s,t.\ s\simeq t \Longleftrightarrow s=_{AC} t,\\
\forall s,t.\ s\rightsquigarrow_{R_\infty} t \Longrightarrow t\prec s.
\end{array}
\]

\subsection{Soundness}

The soundness property of \ACX{} is ensured by the following
invariant:
\begin{thm} 
  For any  configuration $\complbis{E_n}{R_n}$ reachable from
  $\complbis{E_0}{\emptyset}$,
\[
\forall ~s,~t,~~~ (s,t)\in E_n\cup R_n \implies s =_{AC,\X,E_0} t.
\]
\end{thm}

\proof
The invariant obviously holds for the initial configuration and is
preserved by all the inference rules. The rules  
\simplify{}, \compose{}, \collapse{} and \deduce{} preserve the
invariant since for any rule $l\rightarrow r$, if $l =_{AC,\X,E_0} r$,
for any term $s$ rewritten by $\rightsquigarrow_{l\rightarrow r}$ into
$t$, then $s =_{AC,\X,E_0} t$. If \xsolve{} is used to turn an equation
$s\approx t$ into a set of rules $\{p_i \rightarrow v_i\}$, by
definition of $\fsolve$, $p_i = x_i\rho$ and $v_i = t_i\rho$,
where $\fsolveX (\make{s}\approx \make{t}) = \{ x_i \approx t_i\}$ . By
soundness of $\fsolveX$ $x_i =_{\X,\make{s}\approx \make{t}} t_i$. An
equational proof of $x_i =_{\X,\make{s}\approx \make{t}} t_i$ can be
instantiated by $\rho$, yielding an equational proof  $p_i =_{\X,
  s\approx t} v_i$. Since by induction $s=_{AC,\X,E_0} t$ holds, we get
$p_i =_{AC,\X,E_0} v_i$. 
\qed

\subsection{Completeness}

Completeness is established in several steps using a variant of
the technique introduced by Bachmair {\em et al.} in \cite{bachmair86}
for proving completeness of completion. This technique transforms a
proof between two terms which is not under a suitable form into a
smaller one, and the smallest proofs are the desired ones. 

The proofs we are considering are made of elementary steps, either
equational steps, with AC, \X{} and $E_\infty$, or rewriting steps,
with $R_\infty$ and the additional (possibly infinite) rules 
\[
R_\can = \{t \rightarrow \can(t) \mid \can(t) \neq t\}.
\] 
Rewriting steps with
$R_\infty$ can be either $\rightsquigarrow_{R_\infty}$ or
$\rightarrow_{R_\infty}$\footnote{Here,$s\longrightarrow_{R_\infty}t$
  actually means $s\longrightarrow_{AC\backslash R_\infty}t'$ and
  $t=\can_{AC}(t')$. }. 

The measure of a proof is the multiset of the elementary measures of
its elementary steps.  The measure of an elementary step is a
5-tuple of type 

\[
\mathtt{multiset}(\TFX) \times \pNat \times \pNat \times \TFX \times \TFX.
\]

\noindent
It takes into account the number of terms which are in a canonical
form in an elementary proof: the canonical weight of a term $t$,
$w_\can(t)$ is equal to 0 if $\can(t) =_{AC} t$ and to 1
otherwise. Notice that if $w_\can(t)=1$, then $\can(t)\prec t$, and if
$w_\can(t)=0$, then $\can(t)\simeq t$. The measure of an elementary
step between $t_1$ and $t_2$ is defined as follows:
\begin{enumerate}[$\bullet$]
\item When performed thanks to
 an equation, it is equal to $(\multiset{t_1,t_2},\_,\_,\_,\_)$.

\item When performed thanks to a rule $l\rightarrow r\in R_\infty$, it is equal to 
  \[
  (\multiset{t_1},1,w_\can(t_1) + w_\can(t_2),l,r) 
  \quad\mbox{~if~} t_1\rightsquigarrow_{l\rightarrow r}t_2 
  \mbox{~or~} t_1\rightarrow_{l\rightarrow r}t_2 ,
  \] 
  and to 
  \[
  (\multiset{t_2},1,w_\can(t_1) + w_\can(t_2),l,r)
  \quad\mbox{~if~} t_1\leftsquigarrow_{r\leftarrow l}t_2
  \mbox{~or~} t_1\leftarrow_{r\leftarrow l}t_2.
  \] 
  In the case of a $\rightsquigarrow$ step, the measure  is actually 
  $(\multiset{t_i},1,w_\can(t_i),l,r)$ since the reduct is always in a
  canonical form.

\item When performed thanks to a rule of $R_\can$ is equal to 
  \[
  (\multiset{t_1},0,w_\can(t_1) + w_\can(t_2),t_1,t_2)
  \quad\mbox{~if~} t_1\rightarrow_{R_\can}t_2,
  \]
  and to 
  \[
  (\multiset{t_2},0,w_\can(t_1) + w_\can(t_2),t_2,t_1)
  \quad\mbox{~if~} t_1\leftarrow_{R_\can}t_2.
  \] 
\end{enumerate}

\noindent Elementary steps are compared lexicographically using the multiset
extension of $\preceq$ for the first component, the usual ordering
over natural numbers for the components 2 and 3, and $\preceq$ for last ones.
Since $\preceq$ is an AC-reduction ordering, the ordering defined over
proofs is well-founded.

The general methodology is to show that a proof which contains some
unwanted elementary steps can be replaced by a proof with a strictly
smaller measure. Since the ordering over measures is well-founded,
there exists a minimal proof, and such a minimal proof is of the
desired form.

\begin{lem}\label{lemma:acxeq}
A proof containing an elementary step $\longleftrightarrow_{s\approx
  t}$, where $s\approx t\in AC ~\cup~ \X$ is not minimal.
\end{lem}

\proof An elementary equational step using an equation $s\approx t$ of
$AC ~\cup~ \X$ under the context $\mathrm{C}[\_]_p$ can be reduced:
the subproof
\[
\mathrm{C}[s]_p \mathop{\longleftrightarrow}_{s\approx t} \mathrm{C}[t]_p 
\]
is replaced by
\[
\mathrm{C}[s]_p \mathop{\longrightarrow}_{R_\can}^{\{0,1\}} 
\can(\mathrm{C}[s]_p) = \can(\mathrm{C}[t]_p) \mathop{\longleftarrow}^{\{0,1\}}_{R_\can}
\mathrm{C}[t]_p.
\]
The measure strictly decreases, since for the first subproof it
is equal to 
\[
\multiset{(\multiset{\mathrm{C}[s]_p, \mathrm{C}[t]_p},\_,\_,\_,\_)},
\] 
and for the second one, it is equal to
\[
\multiset{(\multiset{\mathrm{C}[s]_p},\_,\_,\_,\_)^{\{0,1\}},
  (\multiset{\mathrm{C}[t]_p},\_,\_,\_,\_)^{\{0,1\}}}.
\]
The rewrite steps $\mathop{\rightarrow}^{\{0,1\}}_{R_\can}$ only
occur on a term which is not AC-equal to a canonical form (which is
denoted by the $\{0,1\}$ exponent). The corresponding elementary
measure occurs in the global measure of the second subproof accordingly.
\qed

\begin{lem}\label{lemma:einftyeq}
A proof containing an elementary step $\longleftrightarrow_{s\approx
  t}$, where $s\approx t\in E_\infty$ is not minimal.
\end{lem}
\proof An elementary equational step using an equation $s\approx t$ of
$E_\infty$ under the context $\mathrm{C}[\_]_p$ can be reduced. Since
$E_\omega$ is empty, there is a completion state where $s\approx t$
disappears, either by \simplify{} or \xsolve.
\begin{enumerate}[$\bullet$]
\item If \simplify{} is used to reduce $s$ into $s'$ by the rule
  $l\rightarrow r$ of $R_\infty$, the subproof
  \[
  \mathrm{C}[s]_p \mathop{\longleftrightarrow}_{s\approx t} \mathrm{C}[t]_p 
  \]
  is replaced by
  \[
  \mathrm{C}[s]_p \mathop{\rightarrow}_{l\rightarrow r} 
  \mathrm{C}[s']_p \mathop{\longleftrightarrow}_{s'\approx t}
  \mathrm{C}[t]_p.
  \]
  The measure strictly decreases, since for the first subproof it
  is equal to 
  \[\multiset{(\multiset{\mathrm{C}[s]_p,
      \mathrm{C}[t]_p},\_,\_,\_,\_)},\] 
  and for the second one, it is equal to
  \[
  \multiset{(\multiset{\mathrm{C}[s]_p},\_,\_,\_,\_),
    (\multiset{\mathrm{C}[s']_p,\mathrm{C}[t]_p},\_,\_,\_,\_)},
  \] 
  and $s \succ s'$.
\item If the rule \xsolve{} turns $s\approx t$ into a set of rules
  $\pi = \{p_i \rightarrow v_i\}$, by definition of $\fsolve$ we have
  $\fsolveX(\make{s}\approx \make{t}) = \{x_i \approx t_i\}$ (denoted
  as $\sigma$) with $p_i = x_i\rho$ and $v_i = t_i\rho$. Since
  $\fsolveX$ is complete, $\make{s}\sigma =_\X
  \make{t}\sigma$. Consider a variable $x$ of $\make{s}$ or
  $\make{t}$,
    \begin{enumerate}[-]
    \item if $x \in \{x_i\}$ then $x\rho\pi = p_i\pi  = v_i$ and
      $x\sigma\rho = t_i\rho = v_i$.
    \item if $x \not\in\{x_i\}$ then $x\rho\pi = x\rho$ (since $x\rho
      \not \in \{p_i\}$) and $x\sigma\rho = x\rho$ (since $x\sigma = x$). 
    \end{enumerate}
    In all cases, $x\rho\pi = x\sigma\rho$. The equational step using
    $s\approx t$ can be recovered as a compound step using $\pi$ and
    $R_\can$ as follows:
    \[
    \begin{array}{l}
      \mathrm{C}[s]_p =
      \mathrm{C}[\make{s}\rho]_p \displaystyle{\mathop{\longrightarrow}^+_{\pi}}\\
      \hspace*{1.5cm}
      \mathrm{C}[\make{s}\rho\pi]_p = 
      \mathrm{C}[\make{s}\sigma \rho]_p \displaystyle{\mathop{\longrightarrow}^{0,1}_{R_\can}} 
      ~~\displaystyle{\mathop{\longleftarrow}^{0,1}_{R_\can}} \mathrm{C}[\make{t}\sigma \rho]_p
      = \mathrm{C}[\make{t}\rho\pi]_p \\[0.4em]
      \hspace*{8cm}
      \displaystyle{\mathop{\longleftarrow}^+_{\pi}} \mathrm{C}[\make{t}\rho]_p  =
      \mathrm{C}[t]_p.
    \end{array}
    \]
    The set of rules $\pi$ belongs to $R_\infty$, and
    the measure of the new subproof is a multiset containing only
    elements of the form $(\multiset{\mathrm{C}[s_i]_p},\_,\_,\_,\_)$,
    where $s_i$ is a reduct of a subterm $s$ or $t$ by an arbitrary number of
    steps of $R_\infty$ and $R_\can$. In any case,
    $\multiset{\mathrm{C}[s_i]_p} \prec \multiset{\mathrm{C}[s]_p,\mathrm{C}[t]_p}$.
    The new subproof is strictly smaller than the measure of the original subproof.
    \qed
  \end{enumerate}

\begin{lem}\label{lemma:rnotsquig}
  A proof containing an elementary rewriting step truly of the form
  $\longrightarrow_{R_\infty}$ or  $\longleftarrow_{R_\infty}$ is not minimal.
\end{lem}
\proof
  Here, each  elementary step $s \mathop{\longrightarrow}_{R_\infty} t$ is
  already a $\mathop{\rightsquigarrow}_{R_\infty}$ step if $t=\can_{AC}(t)$ is in a
  canonical form w.r.t $\can$, or it can be replaced by
  \[
  s \mathop{\rightsquigarrow}_{R_\infty} \can(t)
  \mathop{\longleftarrow}_{R_\can} t.
  \]
  The measure of the first subproof is equal to
  \[
  \multiset{(\multiset{s},1,w_\can(s) +w_\can(t),\_,\_)},
  \]
  and the measure of the second one is equal to
  \[
  \multiset{(\multiset{s},1,w_\can(s),\_,\_),(\multiset{t},0,\_,\_,\_)},
  \]
  with $t\prec s$.  Since $w_\can(t)=1$, the measure strictly
  decreases.

  The case $s \mathop{\longleftarrow}_{R_\infty} t$ is symmetrical.
\qed

\begin{lem}\label{lemma:rnotomega}
  A proof containing an elementary rewriting step of the form
  $\rightsquigarrow_{l\rightarrow r}$ or
  $\leftsquigarrow_{r\leftarrow l}$, where $l\rightarrow r \in
  R_\infty\setminus R_\omega$ is not minimal.
\end{lem}
\proof
An elementary $\rightsquigarrow$ step using a rule $l\rightarrow r$ of
  $R_\infty\setminus R_\omega$ can be reduced. The rule
  $l\rightarrow r$ disappears either by \compose{} or by \collapse.
  \begin{enumerate}[$\bullet$]
  \item If \compose{} reduces $r$ to $r'=\can(r[d])$ by the rule $g\rightarrow
    d$ of $R_\infty$, the subproof
    \[
    \mathrm{C}[l]_p \mathop{\rightsquigarrow}_{l\rightarrow r}
    \can(\mathrm{C}[r]_p)
    \]
    can be replaced by
    \[
    \mathrm{C}[l]_p \mathop{\rightsquigarrow}_{l\rightarrow r'}
    \can(\mathrm{C}[r']_p)  = \can(\mathrm{C}[r[d]]_p) \mathop{\leftsquigarrow}_{d\leftarrow g}
    \mathrm{C}[r]_p.
    \]
    The identity $\can(\mathrm{C}[r']_p)  = \can(\mathrm{C}[r[d]]_p)$
    holds $\mathrm{C}[r']_p$ and $\mathrm{C}[r[d]]_p$ are equal modulo
    $R_\can$, that is $AC~\cup~\X$, and such terms have the same
    canonical forms.
    The measure strictly decreases, since for the first subproof it
    is equal to
    \[
    \multiset{(\multiset{\mathrm{C}[l]_p},1,w_\can(\mathrm{C}[l]_p),l,r)},
    \]
    and for the second one, it is equal to
    \[
    \multiset{(\multiset{\mathrm{C}[l]_p},1,w_\can(\mathrm{C}[l]_p),l,r')),(\multiset{\mathrm{C}[r]_p},0,\_,\_,\_)},
    \]
    with $r' \prec r \prec l$.
  \item If \collapse{} reduces $l$ to $l' = \can(l[d])$ by the rule $g\rightarrow
    d$ in $R_\infty$, the subproof
    \[
    \mathrm{C}[l]_p \mathop{\rightsquigarrow}_{l\rightarrow r}
    \can(\mathrm{C}[r]_p)
    \]
    is replaced by
    \[
    \mathrm{C}[l]_p \mathop{\rightsquigarrow}_{g\rightarrow d} 
    \can(\mathrm{C}[l[d]]_p) = 
    \can(\mathrm{C}[l']_p) \mathop{\longleftarrow}_{R_\can}
    \mathrm{C}[l']_p \mathop{\longleftrightarrow}_{l' \approx r}
    \mathrm{C}[r]_p \mathop{\longrightarrow}_{R_\can}
    \can(\mathrm{C}[r]_p).
    \]
    The measure strictly decreases, since for the first subproof it is
    equal to
    \[
    \multiset{(\multiset{\mathrm{C}[l]_p},1,w_\can(\mathrm{C}[l]_p),l,r)},
    \]
    and for the second one, it is equal to
    \[
    \begin{array}{l}\multiset{
    (\multiset{\mathrm{C}[l]_p},1,w_\can(\mathrm{C}[l]_p),g,d),\\\hspace*{2cm}
    (\multiset{\mathrm{C}[l']_p},\_,\_,\_,\_), 
    (\multiset{\mathrm{C}[l']_p\mathrm{C}[r]_p},\_,\_,\_,\_),
    (\multiset{\mathrm{C}[r]_p},\_,\_,\_,\_),
    }.
    \end{array}
    \]
    The last three elements of the second multiset are
    strictly smaller than the element of the first multiset, since
    $l'\prec l$ and $r \prec l$. The first element of the second
    multiset is strictly smaller than the element of the first
    multiset, since either $g \prec l$, and the fourth component
    decreases, or $g \simeq l$ and $d \prec g$. In this case,
     $l'=d \prec r$. The first four components
    are identical, and the last one decreases.
  \end{enumerate}
The case $\leftsquigarrow$ is symmetrical.
\qed

\begin{lem}\label{lemma:peak1}
A proof containing a peak $s\leftarrow_{R_\can} t
\rightarrow_{R_\can}s'$ is not minimal.
\end{lem}
\proof
  All the terms $s, t$ and $s'$ involved in the peak are equal
  modulo AC and \X, hence $\can(s) = \can(s')$.  The subproof
  \[
  s \leftarrow_{R_\can} t \rightarrow_{R_\can} s'
  \]
  is replaced by 
  \[
  s\rightarrow_{R_\can}^{\{0,1\}} \can(s) = \can(s')
  \leftarrow_{R_\can}^{\{0,1\}}s'.
  \]
    
  \noindent The measure strictly decreases, since for the first subproof it is
  equal to 
  \[
  \multiset{(\multiset{t},0,w_\can(t)+w_\can(s),\_,\_),(\multiset{t},0,w_\can(t)+w_\can(s'),\_,\_)},
  \] 
  and for the second one, it is equal to
  \[
  \multiset{(\multiset{s},0,w_\can(s),\_,\_)^{\{0,1\}},(\multiset{s'},0,w_\can(s),\_,\_)^{\{0,1\}}}.
  \]
  $s$ and $s'$
  are smaller than or equivalent to $t$ ($s,s'\preceq t$), and the
  second component strictly decreases, since $\can(s)$ and
  $\can(s')$ are  in a canonical form and $t$ is not.
\qed

\begin{lem}\label{lemma:peak2}
A proof containing a peak $s\leftsquigarrow_{R_\omega} t
\rightsquigarrow_{R_\omega}s'$ is not minimal.
\end{lem}
\proof
  We make a case analysis over the positions of the reductions.
  
  \begin{enumerate}[$\bullet$] 
  \item In the parallel case, the subproof
    \[
    s \mathop{\leftsquigarrow}_{r\leftarrow l}^p t
    \mathop{\rightsquigarrow}_{g\rightarrow d}^q s'
    \]
    can be seen as 
    \[
    s=\can(t[r]_p[g]_q) \mathop{\longleftarrow}_{R_\can}t[r]_p[g]_q \mathop{\longleftarrow}_{r\leftarrow
      l} t[l]_p[g]_q \mathop{\longrightarrow}_{g\rightarrow d}t[l]_p[d]_q\mathop{\longrightarrow}_{R_\can} \can(t[l]_p[d]_q)=s'.
    \]
      
    The above subproof can be replaced by
    \[
    s=\can(t[r]_p[g]_q) \mathop{\longleftarrow}_{R_\can}^{\{0,1\}}
    t[r]_p[g]_q \mathop{\rightsquigarrow}_{g\rightarrow d}
    \can(t[r]_p[d]_q)\mathop{\leftsquigarrow}_{r\leftarrow l}
    t[l]_p[d]_q \mathop{\longrightarrow}_{R_\can}^{\{0,1\}}
    \can(t[l]_p[d]_q)=s'.
    \]
    The measure strictly decreases, since for the first subproof it is
    equal to 
    \[
    \multiset{(\multiset{t},\_,\_,\_,\_),(\multiset{t},\_,\_,\_,\_)},
    \]
    and for the second one, it is equal to
    \[
    \begin{array}{ll}
      \multiset{(\multiset{t[r]_p[g]_q},\_,\_,\_,\_)^{\{0,1\}},& (\multiset{t[r]_p[g]_q},\_,\_,\_,\_),\\
        & (\multiset{t[l]_p[d]_q},\_,\_,\_,\_),(\multiset{t[l]_p[d]_q},\_,\_,\_,\_)^{\{0,1\}}},
    \end{array}
    \]
    and both terms $t[r]_p[g]_q$ and $t[l]_p[d]_q$ are strictly
    smaller than $t = t[l]_p[g]_q$.

  \item If $q$ is a strict prefix of $p$, this means that
    $l\rightarrow r$ can be used to collapse the rule $g\rightarrow
    d$, which is impossible since the strategy is strongly fair, and
    the application of \collapse{} cannot be infinitely delayed.
  \item The case where $p$ is a strict prefix of $q$ is similar.
  \item If $p$ and $q$ are equal, this means that in both reductions,
    the extended rewriting has been used (second case of
    definition~\ref{def:rewriteac}). Otherwise, again, one rule could
    collapse the other. This means that $l$ and $g$ have the same AC
    top function symbol $u$.  When $l$ and $g$ do not share a common subterm, the
    reasoning is similar to the parallel case.
    Otherwise, if they share a common subterm, since the
    strategy is fair, the head critical pair between $l\rightarrow r$
    and $g\rightarrow d$ has been computed. Let $a^\mu$ the maximal
    common part between $l$ and $g$, $l=_{_{AC}} u(a^\mu,b)$, and
    $g=_{_{AC}} u(a^\mu,b')$. The critical pair is  $u(b',r) \approx
    u(b,d)$. The subterm $t|_p$ where both reductions occur is of the
    form $u(a^\mu,u(b,u(b',c)))$ (or $u(a^\mu,u(b,b'))$ if it
    corresponds exactly to the critical pair).

    The subproof can be replaced by
    \[
    s=\mathop{\longleftarrow}_{R_\can}t[u(u(b',r),c)]_p\mathop{\longleftrightarrow}_{u(b',r)\approx
      u(b,d)} t[u(u(b,d),c]_p\mathop{\longrightarrow}_{R_\can}s'.
    \]
    The measure strictly decreases, since for the first subproof it is
    equal to 
    \[
    \multiset{(\multiset{t},\_,\_,\_,\_),(\multiset{t},\_,\_,\_,\_)},
    \]
    and for the second one, it is equal to
    \[
    \begin{array}{ll}
    \multiset{(\multiset{t[u(u(b',r),c)]_p},\_,\_,\_,\_),\multiset{t[u(u(b',r),c)]_p,t[u(u(b,d),c]_p},\_,\_,\_,\_),\\
    \multiset{t[u(u(b,d),c]_p},\_,\_,\_,\_)},\end{array}
    \]
    and both $t[u(u(b',r),c)]_p$ and $t[u(u(b,d),c]_p$ are
    strictly smaller than $t$.
    \qed
   \end{enumerate}

\begin{lem}\label{lemma:peak3}
A proof containing a peak $s \leftsquigarrow_{R_\omega} t \longrightarrow_{R_\can}s'$ is not minimal.
\end{lem}

The proof of this lemma is partly made by structural induction over
$t$, and we need an auxiliary result in order to study how behave a
proof plugged under a context.

\begin{defi}
Given a context $\mathrm{C}[\bullet]_p$, and an elementary proof
$\CP$, $\CP$ plugged under $\mathrm{C}[\bullet]_p$, denoted
as $\mathrm{C}[\CP]_p$ is defined as follows:
\begin{enumerate}
\item if $\CP$ is an equational step $s\leftrightarrow_{l\approx
    r}t$, $\mathrm{C}[\CP]_p$ is $C[s]_p\leftrightarrow_{l\approx
    r}C[t]_p$,
\item if $\CP$ is a rewriting step $s\longrightarrow_{l\rightarrow
    r}t$, $\mathrm{C}[\CP]_p$ is $C[s]_p\longrightarrow_{l\rightarrow
    r}C[t]_p$,
\item if $\CP$ is a rewriting step $s\rightsquigarrow_{l\rightarrow
    r}t$, $\mathrm{C}[\CP]_p$ is either 
  \[
  C[s]_p\rightsquigarrow_{l\rightarrow r}\can(C[t]_p) \leftarrow_{R_\can}^\Lambda C[t]_p
      \quad \mbox{if $C[t]_p$ is not in a canonical form},\]
 or 
 \[
 C[s]_p\rightsquigarrow_{l\rightarrow r}\can(C[t]_p)
 \quad\mbox{otherwise}.
\]
\end{enumerate}
\end{defi}

This definition is extended to a proof made of several steps, by
plugging elementary each step under the context. Notice that if a proof
$\CP$ relates two terms $s$ and $t$, then $\mathrm{C}[{\mathcal
  P}]_p$ relates $\mathrm{C}[s]_p$ and  $\mathrm{C}[t]_p$. 

\begin{lem}
  Let $\CP_1$ and $\CP_2$ be two proofs which do not contain
  $\rightarrow_{R_\infty}$ nor $\leftarrow_{R_\infty}$. If ${\mathcal
    P}_1$ is strictly smaller than (resp. equivalent to) $\CP_2$,
   then $\mathrm{C}[\CP_1]_p$ is strictly smaller than
  (resp. equivalent to) $\mathrm{C}[\CP_2]_p$. Moreover if
  $\CP_2$ is a step $s\rightsquigarrow_{l\rightarrow r}t$,
  $\mathrm{C}[\CP_1]_p$ is strictly smaller than $C[s]_p\rightsquigarrow_{l\rightarrow r}C[t]_p$.
\end{lem}
\proof
It is enough to show the wanted result for elementary steps.
Let  $\CP_1$ and $\CP_2$ be two elementary steps such that ${\mathcal
    P}_1$ is strictly smaller than $\CP_2$.
\begin{enumerate}[$\bullet$]
\item If $\CP_1$ and $\CP_2$ are $\rightarrow_{R_\can}$ steps, they are of the form
  \[
  s_i \mathop{\longrightarrow}_{R_\can} t_i
  \]
  and the corresponding measures are $(\multiset{s_i},0,w_\can(s_i) + w_\can(t_i), s_i, t_i)$. 
  \begin{enumerate}[-]
  \item if $s_1\prec s_2$, then $C[s_1]_p \prec C[s_2]_p$.
  \item if $s_1\simeq s_2$, and $w_\can(s_1) + w_\can(t_1) <
    w_\can(s_2) + w_\can(t_2)$. Since $s_1\simeq s_2$, by the
    AC-totality of $\preceq$, we know that $s_1=_{AC}s_2$, hence
    $w_\can(s_1) = w_\can(s_2)$. This means that $w_\can(t_1) = 0$ and
    $w_\can(t_2) = 1$. Hence $t_1 =_{AC}\can(t_1)$, $t_1\simeq
    \can(t_1)$ and $t_2 \neq_{AC}\can(t_2)$ and $\can(t_2) \prec t_2$.
    Since $s_1=_{AC} s_2$, $\can(t_1) = \can(t_2)$ holds, hence
    $t_1\prec t_2$.
    
    If we look at the plugged proofs, we have $C[s_1]_p \simeq C[s_2]_p$,
    $w_\can(C[s_1]_p) = w_\can(C[s_2]_p)$, $w_\can(C[t_1]_p) \leq
    w_\can(C[t_2]_p) = 1$ and $C[t_1]_p \prec C[t_2]_p$. The measure is
    even on the first component, and either strictly decreases on the
    second component, or weakly decreases over the four first components,
    and strictly decreases over the last one.  In all cases,
    $\mathrm{C}[\CP_1]_p$ is strictly smaller than $\mathrm{C}[{\mathcal
      P}_2]_p$.

  \item if $s_1\simeq s_2$ and $w_\can(s_1) + w_\can(t_1) = w_\can(s_2)
    + w_\can(t_2)$, this means that $t_1\prec t_2$. The case
    $w_\can(t_1) = w_\can(t_2)=0$ is impossible, since this would imply
    $t_1\simeq \can(t_1) = \can(t_2) \simeq t_2$. Hence $w_\can(t_1) =
    w_\can(t_2)=1$.
    
    If we look at the plugged proofs, we have $C[s_1]_p \simeq C[s_2]_p$,
    $w_\can(C[s_1]_p) = w_\can(C[s_2]_p)$, $w_\can(C[t_1]_p) =
    w_\can(C[t_2]_p) = 1$ and $C[t_1]_p \prec C[t_2]_p$. The measure is
    even on the first four components, 
    and strictly decreases over the last one. 
    $\mathrm{C}[\CP_1]_p$ is strictly smaller than $\mathrm{C}[{\mathcal
      P}_2]_p$.
  \end{enumerate}

\item if $\CP_1$ is a $\rightsquigarrow$-step, and $\CP_2$
  is a $\rightarrow_{R_\can}$ step, necessarily, the first component
  strictly decreases. The measure of $\mathrm{C}[{\mathcal
    P}_1]_p$ is 
  \[
  \multiset{(\multiset{C[s_1]_p},1,w_\can(C[s_1]_p),l_1,r_1),
  (\multiset{C[t_1]_p},0,\_,\_,\_)^{\{0,1\}}},
  \]
  and the measure of $\mathrm{C}[\CP_2]_p$ is
  $(\multiset{C[s_2]_p},0,\_,\_,\_)$, where $t_1\prec s_1 \prec s_2$.
  $\mathrm{C}[\CP_1]_p$ is strictly smaller than
  $\mathrm{C}[\CP_2]_p$.
  
\item if $\CP_1$ is a $\rightarrow_{R_\can}$-step, and $\CP_2$
  is a $\rightsquigarrow$ step, necessarily, the first component
  weakly decreases  and the second component
  strictly decreases.

  The measure of $\mathrm{C}[{\mathcal P}_1]_p$ is
  $(\multiset{C[s_1]_p},0,\_,\_,\_)$ which is strictly smaller than
  the measure of $C[s_2]_p \rightsquigarrow_{l_2\rightarrow r_2}
  C[t_2]_p$, that is
  $\multiset{(\multiset{C[s_2]_p},1,w_\can(C[s_2]_p),l_2,r_2)}$ since
  $s_1 \preceq s_2$.

\item if both $\CP_1$ and $\CP_2$ are $\rightsquigarrow$-steps,
  they are of the form
  \[
  s_i \mathop{\rightsquigarrow}_{l_i\rightsquigarrow r_i} t_i,
  \]
  and the corresponding measures are $(\multiset{s_i},1,w_\can(s_i), l_i, r_i)$. 
  The measure of $\mathrm{C}[{\mathcal P}_1]_p$ is
  \[
    \multiset{(\multiset{C[s_1]_p},1,w_\can(C[s_1]_p),l_1,r_1),
  (\multiset{C[t_1]_p},0,w_\can(C[t_1]_p),C[t_1]_p,\can(C[t_1]_p))^{\{0,1\}}}
  \]
and the measure of $C[s_2]_p
  \rightsquigarrow_{l_2\rightarrow r_2} C[t_2]_p$ is
  $(\multiset{C[s_2]_p},1,w_\can(C[s_2]_p),l_2,r_2)$.

If $s_1\prec s_2$, since $t_1\prec s_1$, $\mathrm{C}[\CP_1]_p$ is strictly smaller than
 $C[s_2]_p \rightsquigarrow_{l_2\rightarrow r_2} C[t_2]_p$.

 Otherwise, $s_1\simeq s_2$ and $s_1=_{AC}s_2$. Hence $w_\can(s_1) =
 w_\can(s_2)$ and the decrease occurs on the last two components.
 Therefore 
\[
\multiset{(\multiset{C[s_1]_p},1,w_\can(C[s_1]_p),l_1,r_1),
 (\multiset{C[t_1]_p},0,w_\can(C[t_1]_p),C[t_1]_p,\can(C[t_1]_p))^{\{0,1\}}}
\] 
 is strictly smaller than 
 \[
 (\multiset{C[s_2]_p},1,w_\can(C[s_2]_p),l_2,r_2).
 \]

\item When a step is an equational step, necessarily the decrease
  occurs on the first component. Since $\prec$ is compatible with
  plugging terms under a context, hence the wanted result.
  \qed
\end{enumerate}

We can now come to the proof of Lemma \ref{lemma:peak3}.
\proof
  
Let us denote by $l\rightarrow r$ the rule of $R_\omega$, and
$g\rightarrow d$ the rule of $R_\can$; since $l$ is in a canonical
form (invariant of the completion run), the reduction using
$g\rightarrow d$ can only take place at a position $q$ which is
above or parallel to the position $p$ of the reduction using
$l\rightarrow r$. We prove by induction that there exists a proof
between $s$ and $s'$ which is strictly smaller than the original
peak.

\begin{enumerate}[$\bullet$]
\item In the parallel case, the subproof
  \[
  s \mathop{\leftsquigarrow}_{r\leftarrow l}^p t
  \mathop{\longrightarrow}_{g\rightarrow d}^q s'
  \]
  can be seen as 
  \[
  \can(t[r]_p[g]_q) \mathop{\longleftarrow}_{R_\can}t[r]_p[g]_q \mathop{\longleftarrow}_{r\leftarrow
    l} t[l]_p[g]_q \mathop{\longrightarrow}_{R_\can} t[l]_p[d]_q.
  \]
  Notice that $t[r]_p[g]_q$ and  $t[r]_p[d]_q$ are equal modulo
  AC,\X, hence have the same canonical form. The above subproof
  can be replaced by
  \[\can(t[r]_p[g]_q)= \can(t[r]_p[d]_q) \mathop{\longleftarrow}_{R_\can}t[r]_p[d]_q \mathop{\longleftarrow}_{r\leftarrow l} t[l]_p[d]_q
  \]
  which is actually
  \[
  s \mathop{\leftsquigarrow}_{r\leftarrow l} s'.
  \]
  
  The measure strictly  decreases, since for the first subproof it is
  equal to 
  \[
  \multiset{(\multiset{t},1,1,l,r),(\multiset{t},\_,\_,\_,\_)},
  \]
  and for the second one, it is equal to
  \[
  \multiset{(\multiset{s'},1,w_\can(s'),l,r)},
  \]
  with $s'\preceq t$.
  
\item In the prefix case, we first prove the wanted result when the
  position $q$ is equal to $\Lambda$. Now we make an induction over
  $p$, in order the establish that there is a proof between $s$ and
  $s'$, with a measure (weakly) smaller than $s\leftsquigarrow
  _{r\leftarrow l}t$, hence strictly smaller than the global measure of
  the peak.  If $p=\Lambda$, rewriting at top with a rule of
  $R_\omega$ is impossible if it is not an extended rewriting, since
  $l$ is in a canonical form. In the extended case, the subproof to be
  replaced has the form
  \[
  \can(u(r,l'))   \mathop{\leftsquigarrow}_{r\leftarrow
    l} t \mathop{\longrightarrow}_{R_\can}^\Lambda s',
  \]
  where $t=_{AC} u(l,l')$, and $s'=\can(u(l,l'))$.  By definition of
  $\can$ and since $l$ is in a canonical form and $u$ is an $AC$
  symbol, $s'$ is AC-equal to $u(l,\can(l'))$. The subproof can be
  replaced by
  \[
  \can(u(r,l')) = \can(u(r,\can(l')))
  \mathop{\leftsquigarrow}_{r\leftarrow l} u(l,\can(l')) =_{AC} s',
  \]
  where the identity $\can(u(r,\can(l'))) = \can(u(r,l'))$ holds since
  $u(r,\can(l'))$ and $u(r,l')$ are equal modulo AC, \X.  The measure
  strictly decreases, since for the first subproof it is equal to
  \[
  \multiset{(\multiset{t},1,w_\can(t),l,r),(\multiset{t},\_,\_,\_,\_)},
  \]
  and for the second one, it is equal to 
  \[
  \multiset{(\multiset{s'},1,w_\can(s'),l,r)},
  \]
  where $s'\prec t$, or $s'\simeq t$ with $w_\can(s') = w_\can(t)$.

  If $p$ is of the form $i\cdot p'$, $t$ is of the form
  $f(t_1,\ldots,t_{i-1},t_i,t_{i+1},\ldots,t_n)$, and the proof to be
  replaced
  \[
  \can(f(t_1,\ldots,t_i[r]_{p'},\ldots,t_n))
  \mathop{\leftsquigarrow}_{r\leftarrow l}
  f(t_1,\ldots,t_i[l]_{p'},\ldots,t_n)
  \mathop{\longrightarrow}_{R_\can}^\Lambda s'.
  \]
  
  We may assume without loss of generality that
  $t_1,\ldots,t_{i-1},t_{i+1},\ldots, t_n$ are in a canonical form,
  since
  \[
  s' = \can(t) =
  \can(f(\can(t_1),\ldots,\can(t_{i-1}),t_i[l]_{p'},\can(t_{i+1})\ldots,\can(t_n)))
  \]
  and
  \[
  \can(f(t_1,...,t_i[r]_{p'},...,t_n)) = 
  \can(f(\can(t_1),...,\can(t_{i-1}),t_i[r]_{p'},\can(t_{i+1})...,\can(t_n))).
  \]
      
  We also denote as 
  \[
  s_0 = f(t_1,\ldots,\can(t_i[r]_{p'}),\ldots,t_n)
  \] 
  and 
  \[
  s'_0 = f(t_1,\ldots,\can(t_i[l]_{p'}),\ldots,t_n).
  \] 
  We know that $\can(t_i[l]_{p'})\preceq t_i[l]_{p'}$, and we
  distinguish between two cases.
  
  \begin{enumerate}[-]
  \item If $\can(t_i[l]_{p'})\prec t_i[l]_{p'}$, then by induction
    hypothesis, there exists a proof $\CP$ between
    $\can(t_i[r]_{p'})$ and $\can(t_i[l]_{p'})$ which is weakly
    smaller than
    \[
    \can(t_i[r]_{p'})\mathop{\leftsquigarrow}_{r\leftarrow l}
    t_i[l]_{p'}.
    \]
    The decreasing is actually strict since an equivalent proof should
    be in one step, and the only possibility is a step of the form
    \[
    \can(t_i[r]_{p'})\mathop{\leftsquigarrow}_{r\leftarrow l}
    \can(t_i[l]_{p'}).
    \]
    However since $\can(t_i[l]_{p'})\prec t_i[l]_{p'}$ and
    $w_\can( t_i[l]_{p'}) = w_\can(t_i[l]_{p'})$ cannot be not
    simultaneously true, such an equivalent step is not possible.
    Among all possible proofs $\CP$, we pick up a minimal
    one. By the previous lemmas, $\CP$ does not contains
    $\rightarrow_{R_\infty}$ steps, hence $f(t_1,\ldots,{\mathcal
      P},\dots,t_n)$ is strictly smaller than
    \[         
    \can(s_0)\mathop{\leftsquigarrow}_{r\leftarrow l} t.
    \]

    If we consider the proof $\CP'$
    \[
    s \mathop{\longleftarrow}_{R_\can}^{\{0,1\}}
    s_0 \xleftrightarrow{f(t_1,\ldots,\CP,\ldots,t_n)}s'_0 
    \mathop{\longrightarrow}_{R_\can}^{\{0,1\}} s',
    \]
    all its elementary steps are strictly smaller than
    $(\multiset{t},1,1,l,r)$. We have seen that this is true for the
    middle part, and also for the left part
    $(\multiset{s_0},0,1,s_0,s)^{\{0,1\}}$, and the right part
    $(\multiset{s'_0},0,1,s'_0,s')^{\{0,1\}}$.
    
    $\CP'$ is a proof between $s$ and $s'$
    which is strictly smaller than $s\leftsquigarrow_{r\leftarrow l}t$.
    
  \item If $\can(t_i[l]_{p'})\simeq t_i[l]_{p'}$, then by the
    AC-totality of $\preceq$, $\can(t_i[l]_{p'})=_{AC} t_i[l]_{p'}$.
    Since $s'=\can(t)$, we know that $s'\preceq t$ and we make a case
    analysis:
    \begin{enumerate}[$*$]
    \item If $s'\simeq t$ then $s'$ is
      actually $\can_{AC}(t)$ which is AC-equal to $t$. $s'$
      contains $t_i[l]_{p'}$ as a subterm and can be reduced
      with $l\rightarrow r$ to $\can(s'[t_i[r]_{p'}])$ which
      is AC-equal to $t[t_i[r]_{p'}]_i$. Hence
      $\can(s'[t_i[r]_{p'}]) = \can(t[t_i[r]_{p'}]_i) = s$
      and the proof
      \[
      s \mathop{\leftsquigarrow}_{r\leftarrow l}s'
      \]
      is equivalent to, hence weakly smaller than $s
      \mathop{\leftsquigarrow}_{r\leftarrow l}t$.
    \item If $s'\prec t$, then we can first see the peak as follows: 
      \[
      s \leftarrow_{R_\can}^{\{0,1\}} s_0 \leftarrow_{r\leftarrow l} t
      \rightarrow_{R_\can} s' = \can(t).
      \]
      We eagerly replace every occurrence of $l$ by $r$ in $s_0$ and
      $s'$, getting respectively $s_1$ and $s''$. Then $s_1$ and
      $s''$ are equal modulo AC and \X, because any proof modulo
      AC and \X{} between $t$ and $s'$ can be replayed by
      replacing the $\sigma$-instances of AC and \X{} used
      originally by $\sigma'$-instances where $x\sigma'$ is $x\sigma$
      where every occurrence of $l$ is replaced by $r$. We get the new proof
      \[
      s \mathop{\longleftarrow}_{R_\can}^{\{0,1\}} 
      s_0 \mathop{\longrightarrow}_{l\rightarrow r}^* 
      s_1 \mathop{\longrightarrow}_{R_\can}^{\{0,1\}}
      \can(s_1)=\can(s'')\mathop{\longleftarrow}_{R_\can}^{\{0,1\}} 
      s'= \can(t).
      \]
      Since $s'\prec t$,
      all terms in the above proof are strictly smaller than $t$,
      hence the measure of this proof is strictly smaller than
      $(\multiset{t},1,1,l,r)$.
    \end{enumerate}

   If the proof occurs under a context $t[\bullet]_q$, we know
    that there is a proof $\CP$ between $s=\can(t[r]_{q\cdot p'})$ and $\can(t)$
    which is weakly smaller than $(\multiset{t[l]_{q\cdot p'}},1,1,l,r)$ (case
    $\rightarrow_{R_\can}$ at $\Lambda$).
    Hence 
    \[
    s \xleftrightarrow{\CP}\can(t)\mathop{\longleftarrow}_{R_\can}^{\{0,1\}} s'
    \]
    is a proof between $s$ and $s'$ which is weakly smaller than
    \[
    \multiset{(\multiset{t[l]_{q\cdot p'}},1,1,l,r), (\multiset{s'},0,1,s',\can(t))^{\{0,1\}}},
    \]
    whereas the measure of the original peak is
    \[
    \multiset{(\multiset{t},1,1,l,r),(\multiset{t},0,2,t,s')}.
    \]
    Since $s'\preceq t$, the measure of the new proof is strictly
    smaller than the measure of the original peak.
    \qed
  \end{enumerate}
\end{enumerate}

\begin{thm}\label{thm:completeness}
If $s$ and $t$ are two terms such that 
\[
s \mathop{\longleftrightarrow}_{AC,\X,E_\infty,R_\infty}^* s',
\]
then 
\[
\can(s)\downsquigarrow_{_R{_\omega}} = \can(t)\downsquigarrow_{_R{_\omega}}.
\]
\end{thm}
\proof
If $s$ and $s'$ are equal modulo
$\mathop{\longleftrightarrow}_{AC,\X,E_\infty,R_\infty}^*$, so are
$\can(s)$ and $\can(s')$.
By the above lemmas, a minimal proof between $\can(s)$ and $\can(s')$ is necessary
of the form 
\[
\can(s) (\rightsquigarrow_{R_\omega}\cup \rightarrow_{R_\can})^*
  (\leftsquigarrow_{R_\omega}\cup \leftarrow_{R_\can})^*
\can(s').
\]
This sequence of steps can also be seen as 
\[
\can(s) \rightarrow_{R_\can}^* (\rightsquigarrow_{R_\omega} \rightarrow_{R_\can}^*)^*
  (\leftarrow_{R_\can}^*\leftsquigarrow_{R_\omega})^* \leftarrow_{R_\can}^*
\can(s').
\]
By definition $\rightarrow_{R_\can}$ cannot follow a
$\rightsquigarrow_{R_\omega}$-step, and $\can(s)$ and $\can(s')$ cannot be
reduced by  $\rightarrow_{R_\can}$, hence the wanted result.
\qed

\subsection{Termination}
The proof of termination partly reuses some facts used for the
termination proof of AC-ground completion (based on Higman's lemma),
but also needs some intermediate lemmas which are specific to our
framework\footnote{We assume that $\bot$ is not encountered,
  otherwise, termination is obvious.}. We shall prove that, under a
strongly fair strategy, $R_\omega$ is finite and obtained in a finite time
(by cases on the head function symbol of the rule's left-hand side),
and then we show that $R_\omega$ will clean up the next configurations
and the completion process eventually halts on
$\complbis{\emptyset}{R_\omega}$.  In order to make our case analysis
on rules, and to prove the needed invariants, we define several sets
of terms (assuming without loss of generality that $E_0 = \can(E_0)$):
\[
\begin{array}{l}
T_0 = \{t \mid \exists t_0,e_1,e_2\in\TFX, e_1\approx e_2 \in E_0
\mbox{~~and~~} t_0 = e_i|_p \mbox{~~and~~} t_0\rightsquigarrow_{R_\infty}^* t\}, \\
T_{0\X} = T_0 \cup \{f_{_\X}(t_1,\ldots,t_n) \mid f_{_\X}\in\Sigma_{\X} \mbox{~~and~~}
\forall i, t_i\in T_{0\X}\},\\
T_1 = \{t \mid t\in T_0 \mbox{~~and~~} \forall p, t|_p \in T_{0\X}\}, \\
T_2 = \{u(t_1,\ldots,t_n) \mid 2 \leq n \mbox{~~and~~} u\in\Sigma_{AC} \mbox{~~and~~}
\forall i, t_i\in T_1\}.
\end{array}
\]

\noindent $T_0$ is the set of all terms and subterms in the original problem as
well as their reducts by $R_\infty$. The set $T_{0\X}$ moreover
contains terms with $\X$-aliens in $T_0$. $T_1$ is the set of terms
that can be introduced by $\X$ from terms of $T_0$ (by solving or
canonizing). $T_2$ is a superset of the terms built by critical pairs.


\begin{lem}\label{term:inv1}
$\forall \gamma, t, s, ~\gamma\in R_\infty\cap T_j^2 \wedge
t\in T_i \wedge t\rightsquigarrow_{\gamma} s
\implies s \in T_i, \mbox{~for~} i,j = 1,2$. \qed
\end{lem}

The proof is by structural induction over terms (for dealing with
rewriting under a context) and by case analysis over $T_i$ when
rewriting at the top level. It uses the (quasi-immediate) fact that
$T_0 \cap T_2 \subseteq T_1$.

\begin{lem}\label{term:inv2}
For all accessible configuration $\complbis{E_n}{R_n}$, 
$E_n\cup R_n \subseteq T_1^2 \cup T_2^2$.
\end{lem}
The proof is by induction over $n$, and uses Lemma~\ref{term:inv1}.

The first step of the termination proof is to show that $R_\omega\cap
T_1^2$ is finite (Lemma~\ref{term:inv4}). It is specific to our
framework, due to the presence of $\X$\footnote{$\X$ may
  change the head function symbol of terms in an equational proof,
  which is not the case of AC in standard ground AC-completion.}.

\begin{lem}\label{term:inv3}
Under a strongly fair strategy, 
if $l\rightarrow r_n$ is created at
  step $n$ in $R_n$ and $l\rightarrow r_m$ at step $m$ in $R_m$, with
  $n < m$, then $r_m$ is a reduct of $r_n$ by $\rightsquigarrow_{R_\infty}$.
\end{lem}
\proof
The proof is by induction over the length of the derivation, and by
case analysis over the rule which has been applied.
\begin{enumerate}[$\bullet$]
\item \xsolve{} applied on $s=t$ cannot create a new rule
  $p\rightarrow v$ with an already present left hand side, because the
  strongly fair strategy implies that $s$ and $t$ are fully reduced, and the
  new left hand side $p$ is a subterm of $s$ or $t$.
\item \simplify{}, \collapse{} and \deduce{} do not create a new rule.
\item \compose{} obviously preserves the invariant.
\qed
\end{enumerate}

\begin{cor}
Under a strongly fair strategy, $R_\infty$ is finitely branching.
\end{cor}
\proof
  If $R_\infty$ is not finitely branching, there exist an infinite
  sequence of rules $(l\rightarrow r_n)_n$ where $l\rightarrow r_n$
  first appears in $\complbis{E_n}{R_n}$. Thanks to
  Lemma~\ref{term:inv3}, since ${R_\infty}$ is included in $\prec$,
  the sequence $(r_n)_n$ is strictly decreasing w.r.t $\prec$. The
  well-foundedness of $\prec$ contradicts the infinity of $(r_n)_n$.
\qed

\begin{lem}\label{term:inv4}
  Under a strongly fair strategy, the set of rules in $\R_{\omega} \cap
  T_1^2$ is finite.
\end{lem}
\proof
  If $l\rightarrow r$ belongs to the set $\R_{\omega}\cap T_1^2$, $l$
  is reduct of a term $l_0$ in $E_0$ by
  $\rsa_{R_\infty}$. Since $\rsa_{R_\infty}$ is
  terminating (because it is included in $\prec$), and finitely
  branching (above corollary), any term has finitely many reducts by
  $\rsa_{R_\infty}$. In particular since $E_0$ is finite, there
  are finitely many possible left-hand side. Moreover since in
  $R_\omega$ two distinct rules have distinct left-hand sides,
  $\R_{\omega}\cap T_1^2$ is finite.
\qed

Here is the second step of the termination proof, finiteness of
$R_{\omega}\cap T_2^2$, which is mostly the same as in the usual
AC-ground completion:

\begin{lem}
  The set of persistent rules in $\R_{\omega}$
  which are in $T_2^2$ is finite.
\end{lem}
\proof
  The set $R_{\omega}\cap T_2^2$ can be divided into a finite union of
  sets, according to the top AC function symbol of the left hand-side
  of the rules. We shall prove that for each $u\in\Sigma_{AC}$, the
  corresponding subset is finite.

  Let $u$ be a fixed AC function symbol, and let $u(l_1,\ldots,l_n)
  \rightarrow r$ be a rule of $R_{\omega}\cap T_2^2$. By definition of
  $T_2$, and by the soundness of $R_\infty$, each $l_i$ is equal modulo
  $ACX,E_0$ to a term $l_i^0$ in $E_0$. Since $l_i$ is irreducible by
  $R_{\omega}$ (otherwise the rule $u(l_1,\ldots,l_n) \rightarrow r$
  would have collapsed), there is a rewriting proof $l_i
  \leftsquigarrow^*_{R_{\omega}}l_i^0$.
  Notice that two distinct rules in $R_{\omega}$ have some distinct
  left-hand sides (otherwise one would have collapsed the other) (this
  implies in particular that $R_{\omega}$ is finitely branching).
  Since $\rightsquigarrow_{R_{\omega}}$ is included in a well-founded
  ordering, and is finitely branching any term has a finite number of
  reducts. Since $E_0$ is finite, each $l_i$ belongs to the {\em
    finite} set of reducts $\red{E_0}$ of $E_0$ by
  $\rightsquigarrow_{R_{\omega}}$.
  By Higman's lemma, if there is an {\em infinite} number of rules
  where the left-hand side is of the form $u(t_1,\ldots,t_n)$, there
  exist two rules $l\rightarrow r$ and $l'\rightarrow r'$, such that
  the multiset of arguments $\multiset{l_1,\ldots,l_n}$ of $l$ is
  included in the multiset of arguments $\multiset{l'_1,\ldots,l'_m}$ of
  $l'$.  This would imply that the second rule collapses by the first
  one, which contradicts its persistence. Hence the wanted result.
\qed

When $R_\omega$ has been proven to be finite, we show that once it is 
obtained, $R_\omega$ will ``clean up'' the configuration within a finite
number of steps, hence the termination:

\begin{thm}
Under a strongly fair strategy, \ACX{} terminates.
\end{thm}
\proof
  When the strategy is strongly fair, $R_{\omega}$ is finite. Moreover
  each rule in $R_{\omega}$ is obtained within a finite number of
  steps. Once all persistent rules are present in the rules of the
  configuration $\complbis{E}{R}$, the rule \xsolve{} always returns
  an empty set of rules. If the measure of a configuration is the
  triple made of the number of remaining critical pairs to generate,
  the multiset of terms in $R$ (compared with $\prec$), and the number
  of equations on $E$, it strictly decreases. 
\qed

\section{Term Abstraction and Multiset Ordering}  
\label{sec:termabstraction}

In this section, we show that a simple preprocessing step allows us to
use a partial multiset ordering instead of a full AC-compatible
reduction ordering in the \ACX{} algorithm. This optimization is
motivated by the fact that although AC-RPO orderings are suitable when
proving termination of completion procedures, they are not easily
implementable in practice. Our preprocessing step is similar to
the \textbf{Ext}ension inference rule found in Abstract Congruence
Closure~\cite{bachmair03jar}.

Let $K$ be a set of constant symbols disjoint from $\Sigma$ and $\CX$
and $\prec_\X$ be a total rewrite ordering on $\CT(\Sigma_X \cup
K)$. We define two sets of terms $\CT_\emptyset$ and $\CT_{AC}$ as
follows:

\[
\begin{array}[h]{lll}
  \CT_\emptyset &=& 
  \left \{~ f(v_1,\cdots,v_n) 
    \quad
    \left| \quad
      \begin{array}[h]{lr}
        f \in \Sigma_\emptyset & \quad\land \\[0.4em]
        arity(f) = n &\land \\[0.4em]
        \bigwedge_{i=1}^n v_i \in \CT(\Sigma_\X \cup K) &\\[0.4em] 
      \end{array}
    \right.
    ~\right\}, \\[3em]
  \CT_{AC} &=& 
  \left \{\quad u(v_1,u(v_2,\ldots,u(v_{n-1},v_n)\ldots))
    \quad 
    \left| \quad
      \begin{array}[h]{lr}
        u \in \Sigma_{AC} & \quad\land \\[0.4em]
        n \geq 2 &\land \\[0.4em]
        \bigwedge_{i=1}^n v_i \in \CT(\Sigma_\X \cup K)
      \end{array}
    \right.
    \quad\right\}.
\end{array}
\]

In order to enable the use of a multiset ordering as an input for
\ACX{}, we have to transform the original set of ground equations
$E$ to a simpler one containing only \textit{abstracted} equations.

\begin{defi}[Abstracted equations]\label{def:abstract} An equation $s \approx t$ is
  said to be abstracted if one of the following statements holds:

\[
\begin{array}[h]{ll}
1.~~ & s,~t \in \CT(\Sigma_\X \cup K),\\[0.5em] 
2. & s \in (\CT_\emptyset \cup \CT_{AC}) ~~\mbox{and}~~ t \in
\CT(\Sigma_\X \cup K),\\[0.5em] 
3. & s,~t \in \CT_{AC} ~~\mbox{and}~~ s(\Lambda) = t(\Lambda).
\end{array}
\]
The set of all abstracted equations is denoted by $\CA$.
\end{defi}

Let $\pi$ be an abstraction function from $\CT_{AC} \cup
\CT_\emptyset$ to $K$ such that if $\pi(s) = \pi(t)$ then $s =_{AC,\X}
t$. Given a set $E^0$ of ground equations, the term abstraction of
$E^0$ consists in applying, as long as possible, the following
inference rules starting from the initial configuration
$\complbis{E^0}{\emptyset}$.

\[
\begin{array}[h]{c}
\inferrule*[left = \textbf{Ab}stract\textbf{1}, Right=$s\approx t \in \CA$]{
\complbis{E \uplus \{s \approx t\} }{E_{\CA}}
}{
  \complbis{E}{E_{\CA} \cup \{s \approx t\} }
}
\\[1em]
\inferrule*[
left={\textbf{Ab}stract\textbf{2}},Right=\mbox{$\CC[f(\vec v)]\approx t\not\in
  \CA$}]
{\complbis{E \cup \CC[f(\vec v)] \approx t}{E_{\CA}}
}{
  \complbis{E \cup \CC[k] \approx t}{E_{\CA} \cup \{f(\vec v) \approx k \}}
} 
\\[1em]
\begin{array}[h]{lll}
\mbox{where,}& & \\[0.3em]
& 1. ~~ & f(\vec v) \in (\CT_{\emptyset} \cup \CT_{AC}) \\[0.3em]
& 2. & k = \pi (f(\vec v)) \\[0.3em]
\end{array}

\end{array}
\]
%


\noindent Propositions \ref{abst:termination} and \ref{abst:correctness}
state, respectively, the termination and the correctness of the
abstraction process.

\begin{prop}
  \label{abst:termination} The application of the rules
  \textbf{Ab}stract\textbf{1} and \textbf{Ab}stract\textbf{2}
  terminates and produces a configuration of the form
  $\complbis{\emptyset}{E_{\CA}^{\infty}}$, where $E_\CA^\infty
  \subseteq \CA$.
\end{prop}
\proof
  The proof of termination is immediate using a decreasing
  measure. The size of a configuration is equal to the total sum of
  the sizes of the terms in its first component. Here, the size of a
  term is recursively defined in a standard way with 1 for the size of
  constants in $K$, and 2 for the size of other constants.
  
  It remains to show that if a configuration is of the form
  $\complbis{E}{E_\CA}$ and $E \neq \emptyset$, at least one rule
  applies. Let $s \approx t$ be an equation in $E$. If $s \approx t
  \in \CA$ the \textbf{Ab}stract\textbf{1} applies. Otherwise, since
  $s\approx t \not\in\CA$, by condition 1. of
  Definition~\ref{def:abstract}, there is a minimal subterm of $s$ or
  $t$ which does not belong to $\CT(\Sigma_\X \cup K)$. This term thus
  has a suitable form to fulfill condition 1. in the rule
  \textbf{Ab}stract\textbf{2} which applies.
 \qed

\begin{prop}
  \label{abst:correctness}
  Let $\complbis{E^0}{\emptyset} \to^*
  \complbis{\emptyset}{E^\infty_\CA}$ be a fixed run of the
  abstraction process. For any terms $s,t \in \CT(\Sigma,\emptyset)$,
  we have:
  \[
  s=_{E^0,\X,AC} t \Longleftrightarrow s=_{E^\infty_\CA,\X,AC} t.
  \]
\end{prop}
\proof
  The direction $\Rightarrow$ is immediate for
  \textbf{Ab}stract\textbf{1}. For \textbf{Ab}stract\textbf{2}, it
  rests on the fact that a step using $\CC[f(\vec v)] \approx t$ can
  be replaced by two steps, the first one using $f(\vec v) \approx k$
  and the second one using $\CC[k] \approx t$.

  In order to prove $\Leftarrow$, we use the following invariant: if
  $\complbis{E}{E_\CA} \rightarrow \complbis{E'}{E'_\CA}$, $s
  =_{E',E'_\CA,\X,AC} t$ and $s$ and $t$ do not contain any constant
  in $K$, then $s =_{E,E_\CA,\X,AC} t$.  This is immediate when the
  rule \textbf{Ab}stract\textbf{1} is applied.  When
  \textbf{Ab}stract\textbf{2} replaces $\CC[f(\vec v)] \approx t$ by
  $\{f(\vec v) \approx k, \CC[k] \approx t\}$, we first replace every
  step using $\CC[k] \approx t$ by a compound step using $\CC[k] \approx
  \CC[f(\vec v)]$ followed by $\CC[f(\vec v)] \approx t$. Then all
  occurrences of $k$ are replaced by $f(\vec v)$ in intermediate terms,
  and the now useless steps using $f(\vec v) \approx f(\vec v)$
  (former $f(\vec v) \approx k$) are removed.  The transformed proof
  is now in $=_{E,E_\CA,\X,AC}$, and since neither $s$ nor $t$ contain
  constants in $K$, they are not affected by these transformations.
\qed

Now that we have shown how to abstract the initial set of equations $E$,
we will define the reduction ordering $\prec$ that we will use in
\ACX{}. We do not need this ordering to be total on the terms in
$\CT(\Sigma_\X\cup K,\emptyset) \cup \CT_{\emptyset} \cup
\CT_{AC}$. We only need a partial reduction ordering which allows us
to get well oriented rewriting rules from the abstracted equations. Let
$\prec_\X^{mset}$ be the multiset extension of $\prec_\X$. Our
reduction ordering is defined by:

\[
\begin{array}[h]{ll}
  1.~~ & \forall v_1,v_2 \in \CT(\Sigma_\X \cup K),~~ v_1
  \prec_\X v_2 \implies v_1 \prec v_2, \\[0.5em]
  2. & \CT(\Sigma_\X \cup K) \prec \CT_\emptyset, \\[0.5em]
  3. & \CT(\Sigma_\X \cup K) \prec \CT_{AC}, \\[0.5em]
  4. & \forall u(\vec v_1), u(\vec v_2) \in \CT_{AC},~~
  \multiset{\vec v_1} \prec_\X^{mset} \multiset{\vec v_2} \implies 
  u(\vec v_1) \prec u(\vec v_2). \\[0.5em]
\end{array}
\]
After that, we have to show that \ACX{} does not introduce
non-abstracted equations when collapsing rules, computing
critical pairs, using canonized rewriting, and solving equations. Hence, the
following lemma:

\begin{lem}
  For any configuration $\complbis{E^\infty_n}{R_n}$ reachable from
  $\complbis{E^\infty_\CA}{\emptyset}$, we have:
  \[
  \forall (s,t) \in (E^\infty_n \cup R_n),\qquad  s \approx t \in \CA.
  \]
\end{lem}
\proof
  The lemma obviously holds for the initial state. For the induction
  step, we can easily show that the abstracted form of equations is preserved by canonized
  rewriting wrt an abstracted rule, hence so as when applying the
  inference rules \textbf{Sim}plify,
  \textbf{Com}pose and \textbf{Col}lapse. Concerning \textbf{Ded}uce,
  we notice by inspecting the
  definition of \fheadcp{}, that when $l \to r$ and $l'\to r'$
  are abstracted oriented equations, so is the resulting critical
  pair. The only subtle case is \textbf{Ori}ent, in particular when
  solving an equation $s\approx t$, with $s\in \CT(\Sigma_\X \cup K)$
  and $t\in \CT_\emptyset\cup \CT_{AC}$. Due to the definition of
  $\prec$ and to the fact that the solver has to fulfill the
  ordering constraints stated in Axiom~\ref{ax:solve}, the solution of
  $s\approx t$ has to be $t\mapsto s$.
\qed

Finally, we notice that $\prec$ is a suitable ordering for the \ACX{} completion procedure since
on the equations in $\CA$, it coincides  with
the AC-RPO ordering based on a precedence $\prec_p$
such that $\Sigma_{\X} \prec_p K \prec_p \Sigma_{\CE} \cup
\Sigma_{AC}$.

\section{Experimental Results}
\label{lab:experiments}

We implemented the \ACX{} algorithm as well as a preprocessing
step that enables the use of a partial multiset reduction ordering
(see Section~\ref{sec:termabstraction}). As described in
Section~\ref{sec:acx}, the state of the procedure is a pair
$\complbis{E}{R}$ of equations and rules. We apply the following
strategy for processing an equality $u\approx v \in E$:
\[
\mathbf{Sim}^*~(
\mathbf{Tri}~ | ~\mathbf{Bot} ~|~ (\mathbf{Ori}~ (\mathbf{Com}~
\mathbf{Col}~ \mathbf{Ded})^*)).
\]
First, $u \approx v$ is simplified as much as possible by
\simplify. Then, if it is not proven to be trivially solved by
\trivial{} or unsolvable by \bottom{}, it is solved by \orient. Each
resulting rule is added to $R$ and then used to \compose{} and
\collapse{} the other rules of $R$. Critical pairs are then computed
by \deduce{}.  


We benchmark \ACX{} and compare its performances with our own SMT
solver \textsc{Alt-Ergo}~\cite{ergo} and some state-of-the-art solvers
(\textsc{Z3} v2.8, \textsc{CVC3} v2.2, \textsc{Simplify} v1.5.4). All
measures are obtained on a laptop running Linux equipped with a
2.58{\it GHz} dual-core Intel processor and with 4{\it Gb} main
memory.  Provers are given a time limit of five minutes for each test
and memory limitation is managed by the system. The results are given
in seconds; we write \TO{} for \emph{timeout} and \OM{} for \emph{out
  of memory}.

Our test suite is made of crafted \emph{ground} formulas which are
valid in the combination of the theory of linear arithmetic
\textsf{LA}, the free theory of equality $\mathcal{E}$ and a small
part of the theory of sets defined by the symbols $\cup$, $\subseteq$,
the singleton constructor $\{\cdot\}$, and the following axioms:

\[
\begin{array}[h]{ll}
  \mathcal{S}_{\cup} & \left\{\begin{array}[h]{rrl}
      \mathrm{\emph{Assoc}}:~&\forall x,y,z. &~x\cup (y \cup z) \approx (x
      \cup y) \cup z\\[0.4em]
      \mathrm{\emph{Commut}}:~&\forall x,y. &~ x\cup y \approx y \cup x
    \end{array}\right.\\[2em]
  \mathcal{S}_{\subseteq}& \left\{\begin{array}[h]{rrl}
      \mathrm{\emph{SubTrans}}:~&\forall x,y,z. &~ x \subseteq y \
      \wedge \ y \subseteq
      z \ \Rightarrow\ x \subseteq z\\ [0.4em]
      \mathrm{\emph{SubSuper}}:~&\forall x,y,z. &~ x \subseteq y \
      \Rightarrow \ x \subseteq y
      \cup z\\[0.4em]
      \mathrm{\emph{SubUnion}}:~&\forall x,y,z. &~ x \subseteq y \
      \Rightarrow \ x \cup z \subseteq
      y \cup z\\ [0.4em]
      \mathrm{\emph{SubRefl}}:~&\forall x. &~ x \subseteq x\\
    \end{array}\right.
\end{array}
\]

\noindent
The theories $\mathcal{E}$ and \textsf{LA} are built-in for all SMT
solvers we use for our experiments. However, contrarily to \ACX{}
which also natively handles associativity and commutativity, SMT
solvers use a generic mechanism for instantiating the axioms
$\CS_\cup$ to reason modulo the AC properties of $\cup$.

In order to get the most accurate information about \ACX{}, we
first benchmark a stand-alone version of our algorithm on ground
formulas that can be proved without $\mathcal{S}_{\subseteq}$. In a
second step, we consider ground formulas that are only provable with
some axioms in $\mathcal{S}_{\subseteq}$. Since these axioms are not
directly handled by \ACX{}, we benchmark a modified version of
\textsc{Alt-Ergo} (to benefit from its instantiation mechanism) with
\ACX{} as its core decision procedure.

In the following, we use the standard mathematical notation
$\bigcup_{i=1}^{d}a_i$ for the terms of the form $a_1 \cup (a_2 \cup
(\cdots \cup a_d))\cdots)$ and we write $\bigcup_{i=1}^{d}a_i; b$ for
terms of the form $a_1 \cup (a_2 \cup (\cdots \cup (a_d \cup
b)))\cdots)$.

\subsection{Benchmark of a stand-alone \texorpdfstring{\textsf{AC($\mathsf{X}$)}}~}

We consider two categories of formulas. The first category $C_1$ is of
the form
\[
\begin{array}{c}
  \bigwedge_{p=1}^n (\{e\} \cup \bigcup_{i=1}^{d} a^p_i) \approx b^p
  ~\implies~
  \underbrace{\begin{array}{l}
   \bigwedge_{p=1}^{n-1} \bigwedge_{q=p+1}^n ~
   \bigcup_{i=d}^{1} a^p_i;b^q \approx
   \bigcup_{i=d}^{1}a^q_i;b^p
 \end{array}}_G,
\end{array}
\]
and the second category $C_2$ is of the form
\[
\begin{array}{c}
  \bigwedge_{p=1}^n (\{t_p- p\} \cup \bigcup_{i=1}^{d} a^p_i) \approx b^p \wedge
  \bigwedge_{p=1}^{n-1} t_p + 1 \approx t_{p+1} 
  \implies G.
\end{array}
\]

Notice that $n$ is the number of hypothesis equations and $d$ is the
maximal depth of AC terms.

\medskip

Proving the validity of $C_1$-formulas only requires the theory
$\mathcal{E}$ and the AC properties of the union symbol. These
formulas are directly provable by \acEmpty{} and the results for this
instance are given in the first column of the table in
Figure~\ref{bench:ac}. In order to prove $C_1$-formulas with SMT
solvers, the axioms in $\mathcal{S}_{\cup}$ have to be put in their
context. The last four columns of the table contain the results for
\textsc{Alt-Ergo}, \textsc{Z3}, \textsc{CVC3} and \textsc{Simplify}.
\begin{figure}[h]
  \begin{center}
    \begin{minipage}[h]{12cm}
      \renewcommand{\arraystretch}{1.1}
      \begin{tabular}[h]{|p{3.5em}|p{5.5em}|p{5.5em}|p{5.5em}|p{5.5em}|p{5.5em}|}
        \hline
        $n$, $d$   & {\sc \acEmpty{}} & 
        {\sc Alt-Ergo} & {\sc Z3} & {\sc CVC3} & {\sc Simplify} \\ \hline
        3, 3   & {\bf 0.01} & 0.19  & 0.22 & 0.40   & 0.18 \\ \hline
        3, 6   & {\bf 0.01} & 32.2  & \OM{} & 132   & \OM{} \\ \hline
        3, 12  & {\bf 0.01} & \TO{} & \OM{} & \OM{} & \OM{} \\ \hline
        6, 3   & {\bf 0.01}         & 11.2  & 1.10  & 13.2  & 2.20 \\ \hline
        6, 6   & {\bf 0.02}         & \TO{} & \OM{} & \OM{} & \OM{} \\ \hline
        6, 12  & {\bf 0.02}         & \TO{} & \OM{} & \OM{} & \OM{} \\ \hline
        12, 3  & {\bf 0.16}         & \TO{} & 5.64  & 242   & 11.5 \\ \hline
        12, 6  & {\bf 0.24}         & \TO{} & \OM{} & \OM{} & \OM{} \\ \hline
        12, 12 & {\bf 0.44}         & \TO{} & \OM{} & \OM{} & \OM{} \\ \hline
      \end{tabular}
      \caption{The results for category $C_1$.}
      \label{bench:ac}
    \end{minipage}
  \end{center}
\end{figure}

In order to prove the validity of $C_2$-formulas, the theory
$\mathcal{E}$, the AC properties of $\cup$ and the theory of linear
arithmetic \textsf{LA} are required. These ground formulas are
directly provable by \acLAMacro{} and the results are given in the
first column of the table in Figure~\ref{bench:arith}. Similarly to
category $C_1$, the last four columns of the table contain the results
for the SMT solvers we considered. Again, the axioms
$\mathcal{S}_{\cup}$ have to be provided in the context, whereas
linear arithmetic is directly handled by the built-in decision
procedures of these provers.
\begin{figure}[h]
  \begin{center}
    \begin{minipage}[h]{12cm}
      \renewcommand{\arraystretch}{1.1}
      \begin{tabular}[h]{|p{3.5em}|p{5.5em}|p{5.5em}|p{5.5em}|p{5.5em}|p{5.5em}|}
        \hline
        $n$, $d$   & {\sc \acLAMacro{}}  & {\sc Alt-Ergo} & {\sc Z3} & {\sc
          CVC3} & {\sc Simplify} \\ \hline 
        3, 3   & {\bf 0.01} & 1.10  & 0.03  & 0.11  & 0.19\\ \hline
        3, 6   & {\bf 0.01} & \TO{} & 3.67  & 4.21  & \OM{}\\ \hline
        3, 12  & {\bf 0.01} & \TO{} & \OM{} & \OM{} & \OM{}\\ \hline
        6, 3   & {\bf 0.02} & 149   & 0.10  & 2.26  & 2.22\\ \hline
        6, 6   & {\bf 0.02} & \TO{} & 17.7  & 99.3  & \OM{}\\ \hline
        6, 12  & {\bf 0.04} & \TO{} & \OM{} & \OM{} & \OM{}\\ \hline
        12, 3  & {\bf 0.27} & \TO{} & 0.35  & 44.5  & 11.2\\ \hline
        12, 6  & {\bf 0.40} & \TO{} & 76.7  & \TO{} & \OM{}\\ \hline
        12, 12 & {\bf 0.72} & \TO{} & \OM{} & \OM{} & \OM{}\\ \hline
      \end{tabular}
      \caption{The results for category $C_2$.}
      \label{bench:arith}
    \end{minipage}
  \end{center}
\end{figure}

\subsection{Benchmark of {\sc Alt-Ergo} with \texorpdfstring{$\mathsf{X}$}~}

We now analyze the performances of \ACX{} when it is used as the
core decision procedure of \textsc{Alt-Ergo}. For that, we consider a
third category $C_3$ of formulas of the form
\[
\begin{array}{c}
  \bigwedge_{p=1}^n \bigcup_{i=1}^{d} \{e^p_i\} \approx b^p 
  \wedge
  \bigcup_{i=1}^{d} \{e + e^p_i\} \approx c^p 
  \wedge
  e \approx 0 
  \implies 
  \bigwedge_{p=1}^n
  c^p
  \subseteq
  (b^p \cup \{e_d^p\}) \cup \{e\}.\\[0.5em]
\end{array}
\]

Proving the validity of $C_3$-formulas requires the theory
$\mathcal{E}$, the AC properties of $\cup$, the theory of linear
arithmetic \textsf{LA} and additionally some axioms in
$\mathcal{S}_{\subseteq}$. We thus \emph{only} provide the axioms
$\mathcal{S}_{\subseteq}$ in the context of the modified version of
\ergo{}, whereas \emph{all} the axioms in $\mathcal{S}_{\subseteq}$
and $\mathcal{S}_{\cup}$ are given in the context of the other SMT
solvers. The results of this category are given in
Figure~\ref{bench:context}.
\begin{figure}[h]
  \begin{center}
    \begin{minipage}[h]{12cm}
      \renewcommand{\arraystretch}{1.1}
      \begin{tabular}[h]{|p{3.5em}|p{5.5em}|p{5.5em}|p{5.5em}|p{5.5em}|p{5.5em}|}
        \hline
        $n$, $d$   & {\sc Alt-Ergo} with \acLAMacro{}   & {\sc Alt-Ergo} &
        {\sc Z3} & {\sc CVC3} & {\sc Simplify} \\ \hline 
        3, 3   & {\bf 0.02} & 3.16  & 0.09  & 10.2  & \OM{}\\ \hline
        3, 6   & {\bf 0.04} & \TO{}& 60.6  & \OM{} & \OM{}\\ \hline
        3, 12  & {\bf 0.12} & \TO{} & \OM{} & \OM{} & \OM{}\\ \hline
        6, 3   & {\bf 0.07} & 188   & 0.18  & 179   & \OM{}\\ \hline
        6, 6   & {\bf 0.12} & \TO{}& \TO{} & \OM{} & \OM{}\\ \hline
        6, 12  & {\bf 0.66} & \TO{} & \OM{} & \OM{} & \OM{}\\ \hline
        12, 3  & {\bf 0.20} & \TO{} & 0.58  & \OM{} & \OM{}\\ \hline
        12, 6  & {\bf 0.43} & \TO{} & \TO{} & \OM{} & \OM{}\\ \hline
        12, 12 & {\bf 1.90} & \TO{} & \OM{} & \OM{} & \OM{}\\ \hline
      \end{tabular}
      \caption{The results for category $C_3$.}
      \label{bench:context}
    \end{minipage}
  \end{center}
\end{figure}

\subsection{Benchmarks analysis}

The results in Figures~\ref{bench:ac}~and~\ref{bench:arith} show that,
contrary to the axiomatic approach, built-in AC reasoning is little
sensitive to the depth $d$ of terms: given a fixed number $n$ of
equations, the running time is proportional to $d$. However, we notice
a slowdown when $n$ increases. This is due to the fact that
\ACX{} has to process a quadratic number of critical pairs
generated from the equations in the hypothesis.  From
Figure~\ref{bench:context}, we remark that \ergo{} with \ACX{}
performs better than the other provers. The main reason is that its
instantiation mechanism is not spoiled by the huge number of
intermediate terms the other provers generate when they instantiate
the AC axioms.

\section{Instantiation Issues}
\label{sec:limit}

Although \ACX{} is effective on ground formulas, its integration
as the core decision procedure of \ergo{} suffers from a {\em bad
  interaction} between the built-in treatment of AC and the axiom
instantiation mechanism of \ergo{} which is roughly done as follows:

\begin{enumerate}[$\bullet$]
\item each axiom of the form $\forall \bar x.\ \mathcal{F}(\bar x)$
  provided in the context comes with a pattern $P$ (also called
  \emph{trigger}) which consists of a set of subterms of $\mathcal{F}$
  that covers~$\bar x$,
\item the solver maintains a set $G$ of \emph{known} terms extracted
  syntactically from the \emph{ground} literals that occur during its
  proof search,
\item $G$ is partitioned into a set of equivalence classes according
  to the ground equalities currently known by the solver,
\item new \emph{ground} formulas $\mathcal{F}\sigma$ are generated by
  matching $P$ against $G$ modulo the equivalence classes.
\end{enumerate}

\noindent
Let us show how this mechanism is used to prove the following ground
formula:
\[
(F_1) \qquad (e \approx d \cup a ~\wedge~ b \subseteq d ~\wedge~ c
\approx a \cup d) ~\Rightarrow~ b \cup a \subseteq c.
\]
For that, we only need to use the \emph{SubUnion} axiom
(defined in Section~\ref{lab:experiments}):
\[\mathrm{\emph{SubUnion}:}\qquad \forall x,y,z.~x \subseteq y \ \Rightarrow \ x \cup z \subseteq y \cup z.
\]
Let us assume that the pattern for this axiom is the term $x \cup z
\subseteq y \cup z$. This pattern is matched against the term $b \cup
a \subseteq c$ by looking for a substitution $\sigma$ such that 
\[
(x \cup z \subseteq y \cup z)\sigma = b \cup a \subseteq c
\]
 modulo the set of equivalence classes 
\[
\{ \{e, d \cup a, a \cup d,c\},\ \{a\},\
\{b\},\ \{d\},\ \{b \cup a\},\ \{b\subseteq d\},\ \{b \cup a \subseteq
c\}\}.
\]
Such a substitution exists and maps $x$ to $b$, $z$ to $a$ and
$y$ to $d$ since the term $c$ is in the same class as $d\cup a$. The
proof of $F_1$ follows from the ground instance $b \subseteq d
\Rightarrow b\cup a \subseteq d \cup a$ of \emph{SubUnion}.

Let us now explain the limitation of the interaction between
\ACX{} and the instantiation mechanism. The hypothesis $e \approx
d \cup a$ is useless (from a logical point of view) to prove $b \cup a
\subseteq c$. Hence, the following formula $F_2$ is equivalent to
$F_1$:
\[
(F_2) \qquad (b \subseteq d ~\wedge~ c \approx a \cup d) ~\Rightarrow~ b \cup a \subseteq c.
\]
However, the cooperation of \ergo{} and \ACX{} fails to prove
$F_2$. The reason is that, since the term $d\cup a$ does not
syntactically occur in $F_2$, the equivalence classes are just 
\[
\{ \{a
\cup d,c\},\ \{a\},\ \{b\},\ \{d\},\ \{b \cup a\},\ \{b\subseteq d\},\
\{b \cup a \subseteq c\}\}
\] 
and the matching algorithm fails to match
$x \cup z \subseteq y \cup z$ against $b \cup a \subseteq c$.

\section{Conclusion }
\label{sec:conclusion}

We have presented a new algorithm \ACX{} which efficiently
combines, in the ground case, the AC theory with a Shostak theory {\X}
and the free theory of equality.  Our combination consists in a tight
embedding of the canonizer and the solver for \X{} in ground
AC-completion. The integration of the canonizer relies on a new
rewriting relation, reminiscent to normalized rewriting, which
interleaves canonization and rewriting rules. We proved the soundness
of \ACX{} by reusing standard proof techniques. Completeness is
established thanks to a proofs' reduction argument, and termination
follows the lines of the proof of ground AC-completion where the
finitely branching result is adapted to account for the theory \X{}.
We showed how a simple preprocessing step allows us to get rid of a
full AC-compatible reduction ordering, and to simply use a partial
multiset extension of a {\em non necessarily AC-compatible} ordering.

\ACX{} has been implemented in the \ergo\ theorem prover.  The
first experiments are very promising and show that a built-in
treatment of AC, in the combination of the free theory of equality and
a Shostak theory, is more efficient than an axiomatic approach for
reasoning modulo AC.

As illustrated in Section~\ref{sec:limit}, the main concern for using
\ACX{} as a core decision procedure in \ergo{} is that it does
not saturate equivalent classes of ground known terms modulo AC. A
naive (and incomplete) solution to this issue would consist in adding,
for each known ground AC-term $t$, a few number of AC equivalent terms
(for instance by bounding the length of the AC equational proof
between them). We rather plan to investigate a more elaborate solution
which would consist in extending the pattern-matching algorithm of
\ergo{} to exploit both ground equalities and properties of AC
symbols.  We also plan to extend \ACX{} to handle the AC theory
with unit or idempotence. This will be a first step towards a decision
procedure for a substantial part of the finite sets theory. Another
future work is the extension of \ACX{} with a user defined first
order rewriting system. This could be achieved by applying our
combination technique to normalized rewriting and normalized
completion~\cite{marche96jsc}.

\section*{Acknowledgment}
\noindent We thank Konstantin Korovin for the
discussion about the ordering used in our implementation, which leads
us to write Section~\ref{sec:termabstraction}. We also thank the
anonymous referees of LPAR-17, TACAS'11 and the LMCS journal for their
remarks which helped us to improve this paper.

\end{document}